\documentstyle[a4paper]{aa}
 \input{psfig.tex}

\newcommand{\Mo}{$\mbox{M}_\odot\;$}

\newcommand{\ms}{$\mbox{millisecond }$}
\newcommand{\lapr}{\raisebox{-.6ex}{\mbox{$\stackrel{<}{\mbox{\scriptsize$\sim$}}\:$}}}

\newcommand{\ApJ}{ApJ}
\newcommand{\ApJS}{ApJS}
\newcommand{\Nat}{Nat}
\newcommand{\Aa}{A \& A}
\newcommand{\AaS}{A \& AS}
\newcommand{\MNRAS}{MNRAS}

\begin{document}

   \thesaurus{03                       
               08.16.6;                
               08.14.1;                
               08.02.3;                
              }

  \title{The X-ray emission properties of millisecond pulsars}
  \author{W.~Becker \and J.~Tr\"umper}
  \offprints{W.~Becker}
 \mail{web@mpe.mpg.de}
  \institute{Max-Planck-Institut f\"ur extraterrestrische Physik,
             D-85740 Garching bei M\"unchen, Germany}
 
  \date{Received: March 6.~1998 / Accepted May 15.~1998}
  \authorrunning{W.~Becker \& J.~Tr\"umper}
  \maketitle

 \begin{abstract} Until now X-radiation from nine millisecond pulsars has been 
  detected. In the present paper we summarize the observations and show the 
  results of a re-analysis of archival ROSAT data. In addition we present the
  results of recent observations of PSR J0437$-$4715 with the ROSAT PSPC 
  and HRI detectors. We show that the pulsed fraction is independent of 
  energy in the range 0.1$-$2.4 keV. The pulse width as measured at X-ray 
  energies is comparable with that observed in the radio domain. An upper 
  limit for the X-ray luminosity of the pulsar's bow-shock visible in
  $H_\alpha$ is found to be $L_x < 2 \times 10^{29}\,\mbox{erg s}^{-1}
  (d/180\mbox{pc})^2$.
  We further report on the discoveries of PSR J1024$-$0719, J1744$-$1134 and 
  J2124$-$3358 with the ROSAT-HRI, making them the first solitary galactic 
  millisecond pulsars detected at X-ray energies. The pulse profile of PSR
  J2124$-$3358 shows marginal evidence for a double peak structure. 
  We finally discuss the observed emission properties of the detected
  millisecond pulsars and conclude that the measured power-law spectra
  and pulse profiles together with the close correlation between the
  pulsar's spin-down energy $\dot{E}$ and the observed X-ray luminosity
  suggests a non-thermal origin for the bulk of the observed X-rays.


\end{abstract}

\section{Introduction}
 The satellite observatories ROSAT, ASCA and EUVE have brought important 
 progress in neutron star and pulsar astronomy. With significantly 
 higher sensitivities compared with previous X-ray satellites they 
 allowed for the first time the detection of X-ray emission from objects as 
 faint as millisecond pulsars.

 Millisecond pulsars form a separate group among the ro\-tation-powered
 pulsars.They are distinguished by their small spin periods $(P\le 20$ ms)
 and high spin stability ($dP/dt\approx 10^{-18} - 10^{-21}$ s/s).
 Consequently, they are very old objects with spin-down ages $P/2{\dot P}$
 of typically $10^9-10^{10}$ years and magnetic dipol components 
 $B_\perp \propto \sqrt{P\;\dot{P}}$ of the order of $10^8 - 10^{10}$ G 
 (see \cite{TaylorManchesterLyneCamilo95}1995).
 More than $\sim 75\%$ of the known disk millisecond pulsars are in 
 binaries with a compact companion star, compared with the $\cong 1\%$ of 
 binary pulsars found in the general population. This gives support 
 to the idea that their fast rotation has been acquired by angular 
 momentum transfer during a past mass accretion phase (\cite{BisnovatyiKoganKomberg74}
 1974; \cite{Alpar82}; \cite{BhattacharyaHeuvel91} 1991; \cite{UrpinGeppertKonenkov98}1998;
 see also the recent results by \cite{WijnandsVanderKlis98} (1998) on SAX 1808.4-3658).
 
 Before the launch of ROSAT, nothing was known on the X-ray emission 
 properties of millisecond pulsars. According to the standard models 
 of cooling neutron stars, millisecond pulsars are too old to expect 
 detectable thermal emission (see \cite{Tsuruta98} 1998 and references therein). 
 Thermal X-ray emission, however, may be emitted from polar caps heated 
 up to temperatures of few million degrees by high energy secondary 
 $e^{\pm}$ streaming back to the neutron star's polar cap regions 
 (\cite{Arons81} 1981; \cite{KundtSchaaf93} 1993; \cite{GilKrawczyk96} 1996).
 Non-thermal processes, which may account for the X-ray emission from 
 millisecond pulsars, are:  magnetospheric emission from relativistic
 particles, characterized by a power-law spectrum;  X-rays from a
 relativistic pulsar wind or from an interaction of that wind with the
 interstellar medium or a close companion star (\cite{AronsTavani93} 1993).
 
 At present, there are 33 rotation-powered pulsars detected in the soft X-ray
 domain (see \cite{BeckerTrumper97} 1997 and references therein)\footnote{\sloppy 
 Updated versions of the Tables 1 and 3 from \cite{BeckerTrumper97} (1997) are
 available online from the URL http://www.xray.mpe.mpg.de/$\sim$web/bt97\_update.html.}. 
 Nine of them belong to the group of millisecond pulsars.
 In this paper we review the recent millisecond pulsar observations at X-ray
 energies and report on results obtained with the ROSAT, EUVE, ASCA and RXTE 
 satellites. The structure of the paper is as follows: After a brief
 description of the analysis techniques in Sect.~\ref{data_anal} we summarize
 the X-ray emission properties of the faint detections (PSR 1957+20, J0751+1807,
 J1012+5307, J1024$-$0719 and J1744$-$1134) in Sect.~\ref{faint}. Results from
 the nearby and bright pulsar PSR J0437$-$4715 are presented in
 Sect.~\ref{0437}. In chapter \ref{1821} and \ref{0218} we summarize recent
 results obtained from the globular cluster pulsar PSR 1821$-$24 and from PSR
 J0218+4232. The discovery of pulsed X-ray emission from the solitary galactic
 millisecond pulsar PSR J2124$-$3358 and the conclusions are presented in
 Sect.~\ref{2124} and Sect.~\ref{con_dis}.

\section{Data analysis\label{data_anal} }

  The ROSAT data presented in this paper were analyzed using the 
  extended scientific analysis software EXSAS (\cite{ZimmermannBeckerBelloni94}1994). 
  Source positions and count rates were obtained from a 
  maximum likelihood analysis of the source photons in combination 
  with a spline fit to the background after removing solar-scattered
  X-rays and the particle background. Count rates were dead-time 
  and vignetting corrected.

  Five of the nine \ms pulsars detected by ROSAT are identified only by
  positional coincidence with the pulsar's radio position (see
  Sect.~\ref{faint}). With about $10-80$ source counts the data of these
  pulsars do not allow a spectral analysis. Two other millisecond pulsars
  (PSR J0218+4232 and J2124-3358) are detected by the ROSAT HRI only, 
  which does not provide spectral information.
  Nevertheless, an estimate for the pulsar's energy flux and luminosity
  within 0.1$-$2.4 keV can be obtained from the observed count rate with
  only few additional assumptions: the pulsar's X-ray spectrum, its distance
  and the absorption column $N_H$.
  Detailed information on the X-ray spectrum is only available for two
  (PSR J0437$-$4715, PSR 1821$-$24) of the nine detected \ms pulsars. In
  both cases the best fitting spectral model implies a power-law nature 
  for the emitted X-radiation.
  \cite{BeckerTrumper97} (1997) have shown recently that with the exception of
  the younger pulsars Crab, B1509$-$58 and the Vela pulsar all pulse-phase
  averaged photon indices measured for rotation-powered pulsars in the soft
  X-ray domain are consistent with a photon-index $\alpha \approx -2$. 
  Calculating luminosities we therefore use the individual photon indices 
  resulting from spectral fits and a {\em canonical} value of $\alpha =-2$ 
  where no spectral information is available.
  The proper motion corrected period derivatives were used to compute $\dot{E}$, 
  $B_\perp$ and the spin-down age $\tau$ for PSR 1957+20 (\cite{CamiloThorsettKulkarni94}1994), 
  PSR J0437$-$4715 (\cite{Bell95}1995; \cite{SandhuBailesManchester97}1997) and PSR 
  J2124$-$3358 (\cite{Bailes97}1997).

  The ROSAT PSPC and HRI temporal resolution of {$130\,\mu$s} and {$64 \,
  \mu$s}, respectively, is sufficient to resolve X-ray pulses from \ms 
  pulsars. 
  A low number of detected source photons does not necessarily preclude a 
  search for pulsation. For instance, {\em evidence} for pulsed X-ray 
  emission was found in the ROSAT all-sky survey data of PSR J0437$-$4715 
  already with a number of $\sim 50$ source counts (\cite{Becker93a}1993a).
  For the reduction of the photon arrival times from the spacecraft
  coordinates and recorded times to the solar system barycenter and
  the barycentric dynamical time-scale (TDB), we performed the standard
  procedures for ROSAT data (see \cite{Becker93b}1993b) using the JPL
  DE200 Earth ephemeris and the pulsar ephemeris valid for the observational  
  epoch.
  Correction of the photon arrival times for a pulsar's binary motion 
  was performed using the method of \cite{BlandfordTeukolsky76} (1976).

  Millisecond pulsars are stable clocks. Given the high precision of the
  pulsar's radio ephemeris, each X-ray photon arrival time can be directly
  related to the pulsar's rotation phase $\phi$ using the relation 
  $\phi_i=\mbox{fractional part of}\,(f\Delta t_i + \frac{1}{2}\,\dot{f}
  \Delta t_i^2)$ with $i=1$ to $N$, where $N$ is the number of photons,
  $f$ and $\dot{f}$ the pulsar frequency and its first time derivative at
  the reference epoch $t_{ref}$ and $\Delta t = t_i - t_{ref}$. The
  statistical significance for pulsations was computed using the 
  $Z^2_n$-test with $n=1$ to $10$ harmonics in combination with the H-Test 
  to determine the optimal number of harmonics in the periodic signal 
  (\cite{DeJager87} 1987; \cite{BuccheriDeJager89} 1989).

  Various methods have been employed to measure the fraction of pulsed
  photons: approaches like fitting sinusoids to the phase histogram 
  using the Rayleigh-Power or estimating the minimum in a light curve 
  to find the DC-level are often valid only for specific pulse shapes 
  (e.g.~real sinusoids) or completely neglect random fluctuations around 
  the lowest level. An additional dependence on the number of bins used 
  for the pulse phase histogram, makes the pulsed fraction often a rather 
  method dependent estimate.
  To avoid these problems we have computed the pulsed fractions by using 
  a bootstrap method. This approach has been recently used by 
  \cite{SwanepoelBeerLoots96}(1996) in the analysis of gamma-ray pulsars. 
  The advantage of this method is that the fraction of pulsed photons is 
  calculated from the data without having to construct a histogram or 
  any other estimate of the light curve.

  Another concern is the number of phase bins used to construct pulse 
  profile histograms for presentation purposes or for comparison with model 
  light curves. Here, the bin width is limited on one side by the instruments
  temporal resolution but on the other side also by the statistical
  significance of the signal. The latter is often not taken into account.
  In this work we have chosen the number of phase bins to construct a pulse 
  profile histogram by using the following approach: denoting the Fourier-power 
  of the i-th harmonic by $R_i$ and taking $m$ as the optimal number of
  harmonics as deduced from the H-test, an exact expression for the optimal
  number of phase bins is given by (De Jager 1998, in prep.)

  \begin{equation}
   M = 2.36 \left( \sum_{i=1}^{m} i^2 R_i^2 \right)^{1/3}
  \end{equation}

\noindent
  This expression compromises between information lost due to binning  
  (i.e.~zero bin width to get all information), and the effect of 
  fluctuations due to finite statistics per bin (i.e.~bin width as large 
  as possible to reduce the statistical error per bin). The total error 
  (bias plus variance) is minimized at a bin width of $1/M$.

\section{The faint detections \label{faint}}

\subsection{The black-widow pulsar: PSR B1957+20}

  The first millisecond pulsar reported in X-rays was the so called
  black-widow pulsar, a 1.6 ms pulsar at a distance of 1.53 kpc, which 
  is in a  close, approximately 9.16h orbit with a low-mass white dwarf 
  companion of $\sim 0.025$ \Mo (\cite{FruchterStinebringTaylor88}1988).
  The source was observed in October 1991 with both the PSPC and the HRI 
  detector. We have reanalyzed the archival ROSAT data (Rev 2) and found 
  the source detected in the PSPC observation with a total of 82 counts of 
  which $\approx 18$ counts belong to the background. With an exposure time 
  of 15\,511s we deduce a PSPC count rate of $(4.2\pm 0.6) \times 10^{-3}$ 
  cts/s in 0.1$-$2.4 keV, approximately twice the number given by Kulkarni
  et al.~(1992) for the range 0.5-2.0 keV. The pulsar is marginally detected 
  in the HRI data, in agreement with the results presented by 
  \cite{FruchterBookbinderGarciaBailyn92}(1992).
  An estimate of the pulsar's X-ray flux from the count rate results in
  $f_x=3 \times 10^{-13}\; \mbox{erg} \mbox{ s}^{-1}\mbox{cm}^{-2}$,
  implying a luminosity of $L_x = 8.5\times 10^{31} (d/1.53\;\mbox{kpc})^2\;
  \mbox{erg s}^{-1}$ and an X-ray efficiency of $L_x/\dot{E} \approx 0.8
  \times 10^{-3}$.
  For the interstellar absorption we have adopted $N_H\approx 4.5\times
  10^{21} \;\mbox{cm}^{-2}$ as estimated by \cite{KulkarniPhinneyEvansHasinger92}(1992)
  from the visual extinction ($A_v \approx 1-2$) of a star close to the
  pulsar position (see \cite{DjorgovskiEvans88} 1988).
 
  To search for a modulation of the detected X-rays at the pulsar's rotation
  period, a photon arrival time analysis was applied to all photons selected 
  from a 35 arcsec aperture centered on the pulsar's X-ray position. The 
  selected counts represent more than 99.9\% of the PSPC source photons with 
  a background contribution of about 15\%.
  Folding the corrected arrival times according to the pulsar ephemeris valid
  for the ROSAT observational epoch did not reveal significant X-ray pulses. 
  A not very restrictive $1\sigma$ upper limit of 60\% has been deduced for 
  the X-ray pulsed fraction. 

  Arons \& Tavani93 (1993) have proposed that X-ray emission from PSR 1957+20
  could arise from the pulsar itself or from the pulsar wind which is
  evaporating the companion star. The low number of counts recorded from this
  pulsar does not allow us to distinguish between the two mechanisms.

\subsection{PSR J0751+1807}

  This binary pulsar has been recently discovered by \cite{LundgrenZepkaCordes95}(1995) 
  in the error box of the unidentified EGRET source CGRO J0749+1807
  (\cite{HartmanBertschFichtel92}1992; \cite{FichtelBertschHartman93}1993) but the 
  association between the pulsar and the putative gamma-ray source could not be 
  established (CGRO J0749+1807 is of marginal significance only and was finally 
  removed from the 2EG and 3EG catalog of gamma-ray sources (\cite{2EG}1995; 
  \cite{HartmanBertschBloom98}1998)).
  The 3.48 ms pulsar and its low-mass white dwarf companion are in a circular
  6.3h orbit. The dispersion measure based pulsar distance is 2 kpc.

  ROSAT observed the pulsar in November 1993 (13\,992s, on-axis) and
  April 1994 (7008s, 37 arcmin off-axis) with the PSPC (\cite{BeckerLundgren96}1996).
  Approximately 50 source counts were
  recorded in the November 93 observation leading to a PSPC count rate of
  ($3.6 \pm 0.6) \times 10^{-3}$ cts/s. The degradation of the PSPC's point
  source sensitivity at 37 arcmin off-axis in combination with a reduced
  exposure time prevented a detection of the pulsar in the April 94 data.
  Estimating the pulsar's X-ray flux and luminosity from the 1993 PSPC count
  rate yields $f_x=8.3 \times 10^{-14}\; \mbox{erg} \mbox{ s}^{-1}\mbox{cm}^{-2}$
  and $L_x = 4 \times 10^{31} (d/2\;\mbox{kpc})^2\;\mbox{erg s}^{-1}$, implying
  an X-ray efficiency of $L_x/\dot{E} \approx 5.3 \times 10^{-3}$.
  Converting the count rate to an energy flux is very sensitive to the
  assumed interstellar absorption. The pulsar's dispersion measure implies
  a value of $9 \times 10^{20}\,\mbox{cm}^{-2}$, making the usual but
  uncertain assumption that there are 10 hydrogen atoms for each free
  electron along the line of sight. The model of \cite{DickeyLockman90} (1990)
  implies $N_H\sim 5 \times 10^{20}\;\mbox{cm}^{-2}$ for the total hydrogen
  column through the Galaxy whereas from the HI survey of \cite{StarkGammieWilson92}(1992)
  we deduce $N_H\sim 4 \times 10^{20}\;\mbox{cm}^{-2}$. The latter is adopted
  for the pulsar's absorption column.

  A timing analysis did not reveal significant X-ray pulses at the expected 
  radio period. A $1\sigma$ upper limit of 70\% is deduced for the fraction 
  of pulsed photons.

\subsection{PSR J1012+5307}

  PSR J1012+5307 is a 5.26 ms pulsar which is in a 14.5h binary orbit with
  a low-mass white dwarf companion. The pulsar was discovered recently during
  a survey for short-period pulsars using the Lovell radio telescope at
  Jodrell Bank (\cite{Luciano95}1995).
  A detection of the companion star at optical wave-length was reported
  from an inspection of the Palomar sky survey plates by the same authors.
  The distance of the pulsar as estimated from its dispersion measure is
  0.52 kpc. 

  A marginal X-ray detection of PSR J1012+5307 was reported recently by 
  \cite{Halpern96} (1996) who found the \ms pulsar $\approx 30'$ off-axis 
  in a serendipitous PSPC observations of approximately 14.5 ksec duration.
  An excess of $\approx 80\pm 20$ counts above the background level 
  within a circle of $2.5'$ and within $1'$ of the pulsar's radio 
  position has been interpreted as a possible detection of the pulsar
  (\cite{Halpern96} 1996).
  Converting the PSPC count rate to an energy flux we find $f_x=4.9  
  \times 10^{-14}\; \mbox{erg} \mbox{ s}^{-1}\mbox{cm}^{-2}$ and $L_x 
  = 1.6 \times 10^{30} (d/0.52\;\mbox{kpc})^2\;\mbox{erg s}^{-1}$ for 
  the X-ray luminosity within 0.1$-$2.4 keV. The latter implies an X-ray 
  efficiency of $L_x/\dot{E} \approx 0.4 \times 10^{-3}$. For the 
  interstellar absorption we used $N_H=7 \times 10^{19}\,\mbox{cm}^{-2}$ 
  as given by \cite{SnowdenHasingerJahoda94}(1994) for the HI hole in Ursa 
  Major.
  No X-ray pulses have been detected from PSR J1012+5307. A $1\sigma$  
  pulsed fraction upper limit is 75\%. A 100 ksec ROSAT HRI observation 
  to confirm the pulsars X-ray detection is scheduled for the second 
  half of 1998.

\subsection{PSR J1024$-$0719 and J1744$-$1134}

 At present there are only nine galactic \ms pulsars which are not in binary
 systems (\cite{Camilo96} 1996). PSR J1024$-$0719 and J1744$-$1134 are two of 
 them. Both pulsars were discovered recently by \cite{Bailes97}(1997) during 
 the Parkes 436 MHz survey of the southern sky.
 PSR J1024$-$0719 has a rotation period of 5.16 ms and a period derivative of 
 $\dot{P}=1.84 \times 10^{-20}\;\mbox{s s}^{-1}$, implying an upper limit 
 to the pulsar age of $\ge 4.4\times 10^9$ years and a rotational 
 energy loss of $\ge 5.3 \times 10^{33}\;\mbox{erg s}^{-1}$. The spin
 parameters of PSR J1744$-$1134 are similar to those of J1024$-$0719: 
 with a period of 4.07 ms and a period derivative of $\;0.86\times 10^{-20}
 \;\mbox{s s}^{-1}$ the pulsars spin-down age is $\ge 7 \times 10^9$ years 
 and the spin-down energy is $\ge 1.42 \times 10^{33}\;\mbox{erg s}^{-1}$. 
 We note that for both pulsars the Shklovskii contribution 
 to $\dot{P}$ is not well constrained so far (see \cite{Bailes97}1997). The  
 spin-down age and the rotational energy loss given above are preliminary
 for that reason. 

 \begin{figure}
  \centerline{\psfig{figure=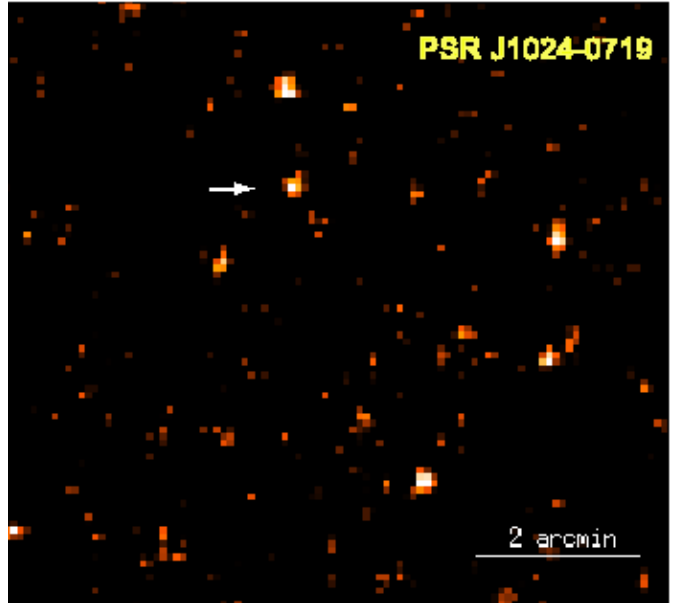,width=8.8cm}}
  \caption[]{Sub-image of the PSR J1024$-$0719 field as observed with the ROSAT
   HRI in November/December 1997. The image has a spatial binning of 5 arcsec.
   The pulsar's X-ray counterpart RX J1024.6$-$0719 is indicated by an arrow.
   Several unidentified X-ray sources are detected in the pulsar's neighborhood,
   only few arcmin distant from the pulsar \label{1024_eps}}
 \end{figure}
 
 Both millisecond pulsars are at a relative close distance to us. For PSR 
 J1024$-$0719 the Taylor \& Cordes model implies $d \sim 350$ pc based on the 
 pulsar's dispersion measure (\cite{Bailes97}1997), whereas for PSR J1744$-$1134 
 the radio parallax has been measured recently and yielded a distance of about 
 $260 \pm 60$ pc (Bailes 1997, priv.~comm.).

 ROSAT HRI observations of PSR J1024$-$0719 and J1744$-$1134 were performed
 in November/December and September 1997 for an effective exposure time of 
 80\,635 sec and 61\,238 sec, respectively. The X-ray sources RX J1024.6$-$0719
 (RA=10:24:38.60, DEC=$-$07:19:21.5) and RX J1744.4$-$1134 (RA=17:44:29.35, 
 DEC=$-$11:34:52.2) were detected with a significance of $\sim 4-5\sigma$, 
 only $2-3$ arcsec distant from the pulsars' positions (\cite{BeckerTrumper98}
 1998; \cite{BeckerTrumperHasinger98}1998, see Fig.\ref{1024_eps}). The 
 positional offset between RX J1024.6$-$0719, RX J1744.4$-$1134 and the pulsars' 
 radio position is well within the uncertainty of the satellite pointing
 accuracy. The probability that RX J1024.6$-$0719 and RX J1744.4$-$1134 
 are chance superpositions of unrelated background object is $\sim 10^{-4}$ 
 and $\sim 3 \times 10^{-5}$, respectively, estimated from the density of 
 X-ray sources in the HRI field of view. We therefore accept RX J1024.6$-$0719 
 and RX J1744.4$-$1134 as the likely X-ray counterparts of the millisecond 
 pulsars PSR J1024$-$0719 and J1744$-$1134.

 Converting the HRI source count rate of $(3.3\pm 0.9)\times 10^{-4}$ and
 $(2.9\pm 0.9)\times 10^{-4}$, as deduced for J1024$-$0719 and J1744$-$1134,
 to an energy flux we find $f_x\sim 2 \times 10^{-14}\; \mbox{erg}
 \mbox{ s}^{-1} \mbox{cm}^{-2}$ within 0.1$-$2.4 keV for both sources. The
 latter implies an X-ray luminosity of $L_x = 3\times 10^{29}\,(d/0.35\,
 \mbox{kpc})^2\,\mbox{erg s}^{-1}$ and $L_x = 2 \times 10^{29}\,(d/0.26\,
 \mbox{kpc})^2\,\mbox{erg s}^{-1}$ for J1024$-$0719 and J1744$-$1134. For
 the interstellar absorption we used $N_H=2\times 10^{20}\,\mbox{cm}^{-2}$
 and $10^{20}\,\mbox{cm}^{-2}$ respectively, as deduced from the pulsars'
 dispersion measure. The number of recorded counts was not sufficient to
 perform a search for pulsations.

\newcommand{\PSR}{PSR J0437$-$4715 }

\section{The nearby and bright pulsar PSR J0437-4715 \label{0437}}

  The first millisecond pulsar for which a more detailed knowledge of its 
  X-ray emission properties has become available is the 5.75 ms pulsar PSR 
  J0437$-$4715. 
  The source, which was discovered in the Parkes southern sky survey by 
  \cite{JohnstonLorimerHarrison93}(1993) and which is in a close 5.74-day circular orbit 
  around a $\approx 0.25$ \Mo white-dwarf companion (\cite{JohnstonLorimerHarrison93}1993;
  \cite{Becker93a}1993a; \cite{Bailyn93} 1993; \cite{DanzigerBaadeDellaValle93}1993) 
  is the nearest and brightest \ms pulsar known. 
  Recent measurements of the annual parallax have adjusted the pulsar's 
  distance to be $178\pm 26$ pc (\cite{SandhuBailesManchester97}1997), placing 
  it somewhat further away than originally indicated by the dispersion measure 
  and the electron density model of \cite{TaylorCordes93} (1993). The parallax 
  distance further constrains the pulsar's proper motion and revises its intrinsic
  period derivative down to $\dot{P}_{int}=(0.8\pm 0.7)\times 10^{-20}$, 
  approximately a factor of 7 smaller than the observed one $\dot{P}_{obs}=
  5.73\times 10^{-20}$ (\cite{Bell97}1997). Based on $\dot{P}_{int}$ the 
  pulsar's spin-down energy and polar magnetic field is found to be $\log 
  \dot{E}= 33.22^{+0.3}_{-0.9}$ erg/s and $\log B_\perp=8.34^{+0.13}_{-0.44}$ 
  G, respectively. The pulsar's characteristic age is $ P/2\dot{P}_{int}\;
  \lapr\;6\times 10^9\,\mbox{yr}$.

  X-ray emission from the pulsar was first detected in the ROSAT all-sky 
  survey by \cite{Becker93a}(1993a). Based on a serendipitous $\approx 6$ ksec 
  pointed observation with the ROSAT PSPC \cite{BeckerTrumper93} (1993) found that 
  the X-ray emission is pulsed and the pulsar's soft X-ray spectrum is best
  described by a single power-law. An energy dependent pulsed fraction showing
  a peak at $\sim 0.9$ keV was found by fitting a sine wave to the X-ray pulse 
  profiles obtained for different energy ranges. Because it appeared difficult 
  to explain this apparent energy dependence in terms of single-component 
  emission \cite{BeckerTrumper93} (1993) proposed that the radiation from PSR 
  0437$-$4715 is a mixture of spin-modulated thermal emission from heated polar 
  caps combined with unpulsed emission from a pulsar wind or plerion.
  A reanalysis of the serendipitous ROSAT data together with data taken with the
  EUVE satellite at $65-120$ \AA\/ (\cite{HalpernMartinMarshall96}1996) confirmed
  the results of \cite{BeckerTrumper93} (1993) but has weakened the case of an
  energy dependence of the pulsed fraction by using improved pulsar ephemeris
  and a different method to estimate the fraction of pulsed photons.

  In the following we present the results from a re-analysis of the archival
  ROSAT data of PSR 0437$-$4715 including unpublished data from subsequent
  observations with both the ROSAT PSPC and HRI detectors. 

\subsection{Observations \label{0437_obs}}

\begin{figure}
  \centerline{\psfig{figure=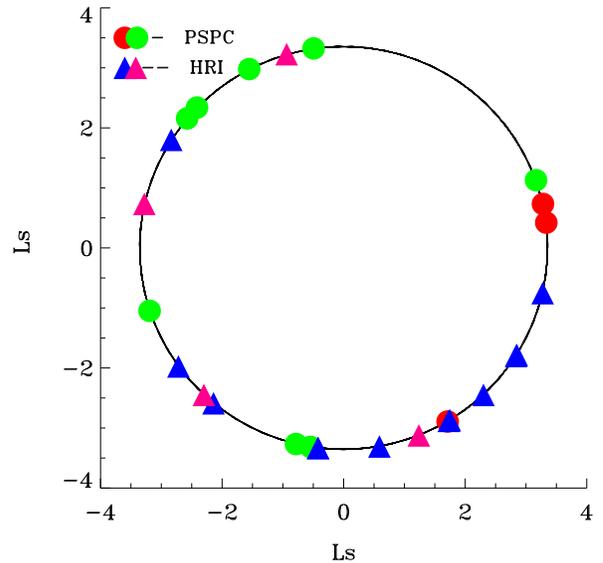,width=8.8cm}}
  \caption[]{Coverage of the pulsar's binary orbit by X-ray observations using
  ROSAT. Circles indicate observations with the PSPC in the focus of the
  telescope whereas triangles represent HRI observations.} \label{0437_ros_orb}
 \end{figure}

  Since its discovery in 1993, \PSR has been the target of several pointings
  with both the PSPC and the HRI. A total of $\sim 20$ ksec PSPC observations
  and $\sim 60$ ksec HRI observations have been performed with different
  sensitivities. The coverage of the pulsar's binary orbit by ROSAT
  observations is illustrated in Fig.~\ref{0437_ros_orb}.
 
  Pointed observations with the ROSAT PSPC were performed on 1992, September 
  20$-$21 for an effective exposure time of 5846 s and on 1994 July 2$-$4 for
  9885 s and 4459 s, respectively. The latter observation was done with the
  boron filter. As a consequence, the pulsar's PSPC count rate is reduced to
  $0.042\pm 0.003$ cts/s, compared with a count rate of $0.204 \pm 0.005$
  cts/s for the filter-off observation.
  In addition, ROSAT observations with the HRI were performed between 1994 July 
  18th and August 24th and on 1997 September 22$-$26 for an effective exposure 
  time of 35\,209 and 22\,928 seconds, respectively. In the later observation 
  the satellite wobble-mode was disabled. The pulsar's HRI source count rate 
  is $0.042\pm 0.002$ cts/s.

\subsection{Spatial emission properties of PSR J0437$-$4715 \label{0437_spatial}}

  \PSR is one of few \ms pulsars for which an $H_\alpha$ bow-shock nebula has
  been detected (\cite{Bell93}1993;~1995). Others are the black-widow pulsar PSR
  1957+20 (\cite{KulkarniHester89} 1989) and PSR J2225+6535 which is surrounded by 
  the so called Guitar Nebula (\cite{CordesRomaniLundgren93}1993; 
  \cite{RomaniCordesYadigaroglu97}1997).
  It is generally believed that the pulsar wind, a mixture of charged particles and 
  electromagnetic radiation, shocks the interstellar medium (see \cite{AronsTavani93} 
  1993) giving rise to a bow-shock. Such a shock region can produce both thermal 
  emission and synchrotron radiation. Only the local conditions, such as the magnetic 
  field and particle density, determines if X-rays are produced and which emission 
  process dominates.

\begin{figure}
  \begin{picture}(88,88)(0,2)
   \put(0,0){{\psfig{figure=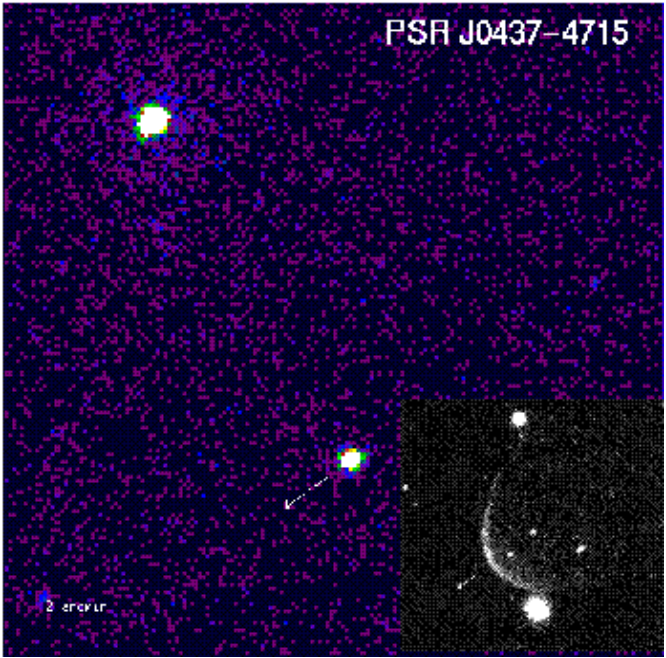,width=8.8cm}}}
  \end{picture}
  \vspace{0ex}
   \caption[]{Sub-image of the pulsar field as observed with the ROSAT HRI in
   September 1997 (satellite wobble disabled). The image has a spatial
   binning of 2.5 arcsec.
   The 4.3 arcmin neighboring source RX J0438.1-4710, identified to be a 
   Seyfert 1 Galaxy at $z=0.051$ (\cite{Grupe96} 1996) is visible to the upper left. 
   The inset in the lower right corner shows the pulsar's optical bow-shock
   nebula as observed by A.~Fruchter in $H_\alpha$. The arrow
   indicates the pulsars proper motion direction, which has an orientation of
   $\sim 122^\circ$ measured anticlockwise from north. \label{0437_hri} \\[-4ex] }
\end{figure}

  The optical bow-shock nebula of PSR J0437-4715 has a standoff distance from the
  pulsar position of about 12 arcsec (see Fig.\ref{0437_hri}). Diffuse optical
  emission of the nebula is visible up to about 1 arcmin from the pulsar position 
  along the side wings of the bow-shock hyperboloid. The pulsar's proper motion 
  direction is $122^\circ \pm 2^\circ$ (\cite{Bell95}1995), measured anticlockwise 
  from north.

\begin{figure}
  \begin{picture}(90,92)(0,18)
  \put(0,0){{\psfig{figure=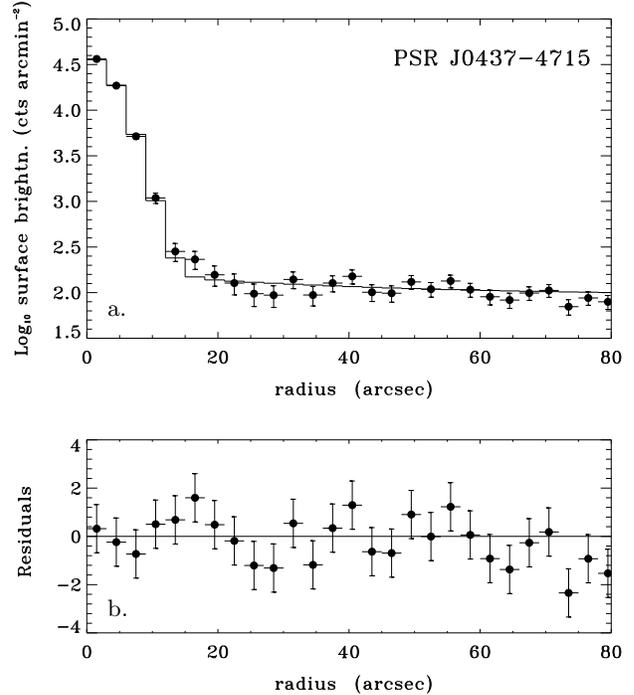,width=8.8cm}}}
  \put(17,68){a.}
  \put(17,28){b.}
  \end{picture}
   \caption[]{{\bf a.}$\,$ The radial surface brightness distribution within 
    an 80 arcsec circle centered on the pulsar PSR J0437$-$4715 and based on the 
    1997 HRI data. The histogram incorporates a point source and an additional 
    broadening of the PSF to account for an uncorrected 2.6 arcsec attitude 
    error, smearing out the model point spread function. {\bf b.}$\,$ Residuals
    of the PSF fit in units of $\sigma$. The fit has a reduced $\chi^2$ of 0.87.
    \label{0437_psf} \\[-4ex] }
\end{figure}

  With a spatial resolution of $\sim 25$ arcsec (FWHM) \PSR appears to be point 
  like in the available PSPC data. A source extension at energies below $\sim 0.2$ 
  keV is found to be caused by the well known ghost imaging effect, an instrumental 
  artefact which mimics an extended source for ultra-soft photons.
  Compared with the ROSAT PSPC the spatial resolution of the HRI detector is higher
  by about a factor of $\sim 5$. The HRI therefore is more suitable to search for an 
  extended emission component on small angular scales.
  Inspecting the 1994 HRI data we found a small elongation of the pulsar's X-ray 
  counterpart towards a direction of $105 \pm 5^\circ$ measured anticlockwise
  from  north. Since this orientation is consistent with the satellites wobble 
  direction ($\sim 102^\circ$) we conclude that this elongation is an artefact 
  of uncorrected residual wobble rather than a feature associated with the 
  pulsar's $H_\alpha$ nebula. 
  To prevent such a source extension due to incomplete attitude correction 
  we observed the pulsar in 1997 with the satellite wobble disabled.
  Figure \ref{0437_hri} shows the HRI image of the pulsar field based 
  on the 1997 data. As can be seen from this image there is no indication 
  of an ellipsoidal elongation in the orientation of the optical bow-shock 
  nebula as found in the 1994 HRI data. Figure \ref{0437_psf} shows that the 
  source extent of the pulsar's X-ray counterpart is in agreement with the 
  ROSAT HRI point-spread-function (see also \cite{Boese98} 1998).

\begin{figure}
  \begin{picture}(88,48)(0,-2)
  \put(0,0){\centerline{{\psfig{figure=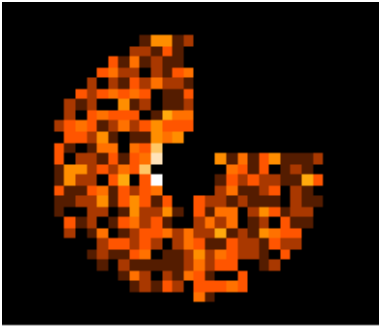,width=5cm}}}}
  \end{picture}
 \vspace{-4ex}
  \caption[]{Sky region around \PSR selected to quantify a possible X-ray emission
  component from the side wings of the pulsar's bow-shock nebula. The mask has an
  inner radius of $12"$ and and outer radius of $60"$ in the orientation of the
  optical image of the pulsar bow-shock. The pixel size is 5x5 arcsec. Based on
  the merged 1994 and 1997 HRI data we find in total 785 counts within the covered
  area.} \label{0437_mask}
 \end{figure}

  To search for a possible X-ray emission component from the side wings of the 
  bow-shock nebula we used the  merged 1994 and 1997 HRI data. 785 counts were
  selected from within three quarter of a ring around PSR J0437$-$4715, having an 
  inner radius of $12"$ and and outer radius of $60"$ in the orientation of the 
  optical image of the pulsar bow-shock. 
  Figure \ref{0437_mask} shows the selected sky region scaled up by a factor 6 
  compared with Fig.~\ref{0437_hri}. Its area is approximately $2.6\,\mbox{arcmin}^2$, 
  implying a surface brightness of $\sim 302\,\mbox{cts arcmin}^{-2}$.
  Shifting the mask to source free regions in the neighborhood of PSR J0437-4715 
  we find that on average 642 of the 785 counts in the mask belong to the background.  
  The excess of 143 counts above the background level is about 6\% of the $\sim 
  2480$ pulsar counts falling within the 12 arcsec inner radius of the mask and 
  thus is well explained by mirror scattering (\cite{DavidHarndenKearnsZombeck97}1997).
  For the $3\sigma$ count rate upper limit of a diffuse nebula component we find
  $3\times \sqrt{785}/58\,138\;\mbox{cts/s} = 1.4 \times 10^{-3}$ cts/s
  (see \cite{Becker95} 1995, p65).
  Assuming a power-law spectrum with a photon-index $\alpha= -2$ and a column density 
  of $N_H=0.8\times 10^{20}\,\mbox{cm}^{-2}$ (see Sect.~\ref{0437_spec}) the count 
  rate upper limit corresponds a $3\sigma$ flux upper limit of $< 5 \times 10^{-14}
  \,\mbox{erg s}^{-1}\,\mbox{cm}^{-2}$ and a luminosity of $ < 2\times 10^{29}\,
  \mbox{erg s}^{-1}\, (d/180\mbox{pc})^2$ in 0.1$-$2.4 keV. Compared with the pulsar's 
  spin-down energy this is less than $0.2\times 10^{-3}\,\dot{E}$, similar to the 
  $3\sigma$ conversion upper limit found for a possible bow-shock contribution to the
  X-ray emission from 1957+20 (\cite{FruchterBookbinderGarciaBailyn92}1992).
   
\subsection{Spectral emission properties \label{0437_spec} }

\begin{figure}
  \begin{picture}(85,85)(0,-6)
  \put(0,0){\psfig{figure=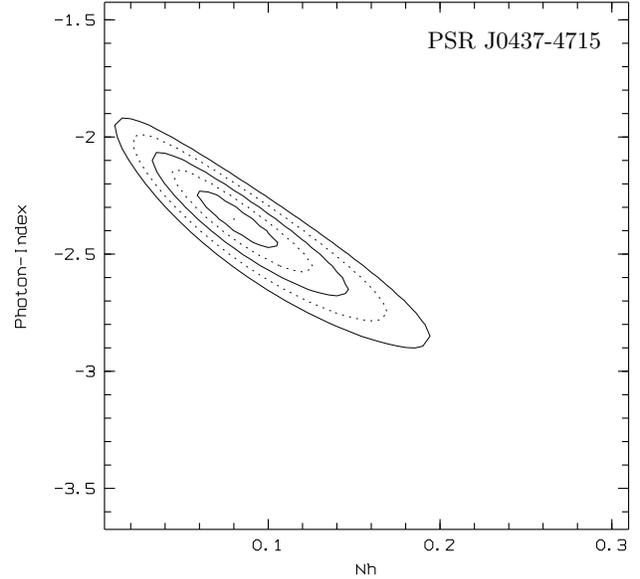,width=8.2cm,clip=}}
  \put(55,70){PSR J0437-4715}
  \end{picture}
  \vspace{-3ex}
  \caption[]{Contour plot of $\chi^2$ as a function of the Galactic
  column density $N_H \times 10^{21}\,\mbox{cm}^{-2}$ and the photon-index
  $\alpha$. The confidence levels range from $1\sigma$ to 5$igma$.}
  \label{0437_pwl_grid}
 \end{figure}

  Becker \& Tr\"umper (1993) have shown that the pulsar's soft X-ray spectrum 
  in 0.1$-$2.4 keV is best described by a power-law 
  $dN/dE\propto E^{\alpha}$. Figure \ref{0437_pwl_grid} shows the $\chi^2$ 
  contour plot of this model in  the $N_H - \alpha$ plane. The best fitting 
  parameters for this model, deduced from the merged 1992 and 1994 PSPC 
  data, are a column density $N_H=(8\pm 2.5)\times 10^{19}\,\mbox{cm}^{-2}$ and 
  a photon-index of $\alpha=-2.35\pm 0.15$. This is in agreement with the results 
  found by \cite{HalpernMartinMarshall96}(1996) based on a combined EUVE/ROSAT spectral 
  analysis using the Sep.~1992 data only. The 1994 PSPC data taken with the Boron 
  filter\footnote{The advantage of the Boron filter is to improve the PSPC's 
  spectral resolution in the carbon band ($0.1-0.288$ keV) by a further 
  subdivision of this energy range into two bands. See 
  \cite{StephanSchmittSnowdenMaierFrischke91}1991 
  for informations on the filter transmission.} were used for a cross check 
  of the fitted power-law spectral parameters by computing the count rate ratio 
  for an observation with and without a Boron filter and comparing this with 
  the measured value. The result was found to be in agreement with the filter-off 
  spectral fittings. 

  The pulsar's energy flux deduced from the best fitting power-law model is $f_x=
  1.9 \pm 0.2 \times 10^{-12}\,\mbox{erg s}^{-1}\mbox{cm}^{-2}$ in 0.1$-$2.4 keV, 
  implying an X-ray luminosity of $L_x=7.3 \pm 0.8 \times 10^{30}\,\mbox{erg s}^{-1}\, 
  (d/180 \mbox{pc})^2$ and a conversion factor of $L_x/\dot{E} \cong 4\times 10^{-3}$.

  \begin{figure}
  \begin{picture}(88,140)(0,10)
  \put(0,75){\psfig{figure=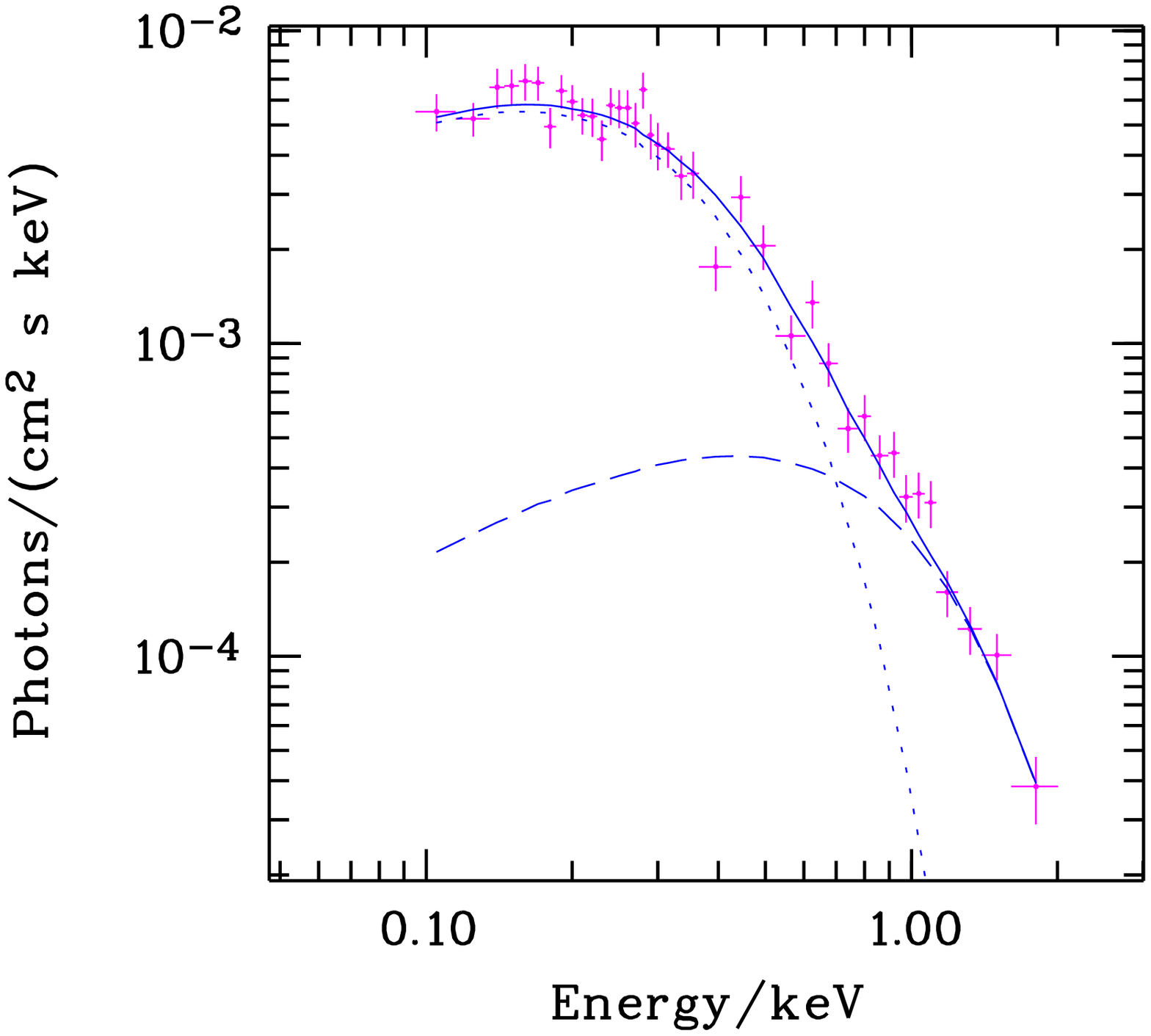,width=8.8cm,clip=}}
  \put(0,0){\psfig{figure=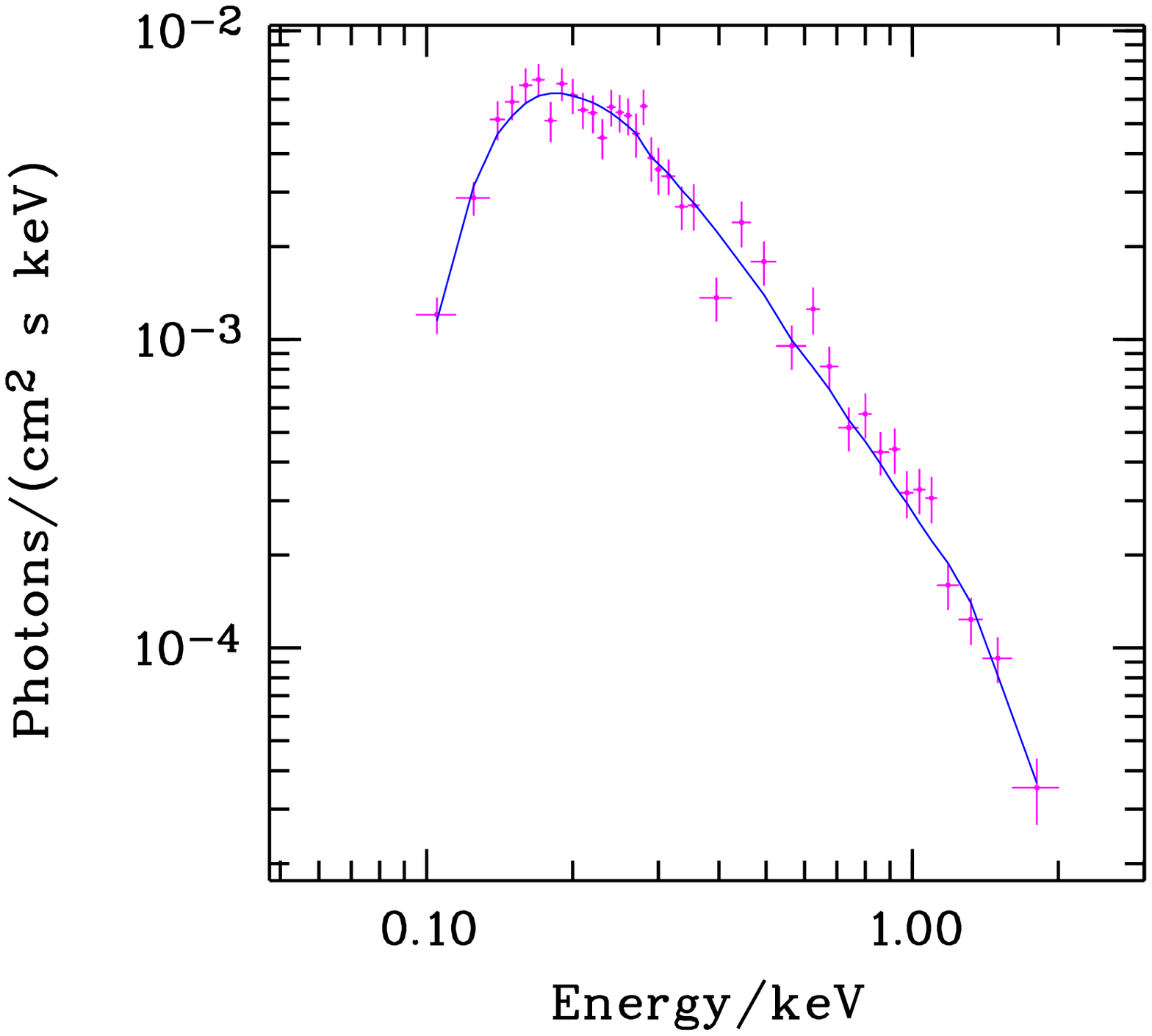,width=8.8cm,clip=}}
  \put(25,94){a.}
  \put(25,18){b.}
  \end{picture}
  \vspace{2ex}
  \caption[]{{\bf a.}$\,$The ROSAT PSPC data of PSR J0437$-$4715 fitted with a two 
   component black body model. The harder component (dashed line) represents the 
   black-body spectrum as deduced from spectral fits to ASCA data (see text). The dotted 
   line represents the soft component as obtained from a spectral fit to the ROSAT 
   data. {\bf b.}$\,$ The PSPC data fitted with a broken power-law. The break energy 
   in this model is 1.4 keV. The slopes of the first and second power-law components 
   are $-2.35$ and $-4.5$, respectively.} \label{0437_brpl_asca} \label{0437_bb_asca}
 \end{figure}  
               
  It was already shown by \cite{BeckerTrumper93} (1993) that a single black-body 
  spectrum does not fit the ROSAT data and leaves a residual hard excess above 
  $\sim 0.5$ keV (see also \cite{Becker95} 1995, p125), whereas a double black-body 
  model yields acceptable spectral fits. 
  Based on ASCA observations, \cite{KawaiTamuraSaito98}(1998 -- hereafter KTS98) have 
  reported recently that the pulsar spectrum beyond 0.7 keV fits better with a black-body 
  rather than with a composite (power-law plus black-body) or simple power-law spectrum.
  The authors deduced a temperature of $T \sim (3 \pm 0.5) \times 10^6$ K, a black body
  radius of $R_{bb}\sim 54\,\mbox{m}\,(d/180\,\mbox{pc})$ and a bolometric flux of
  $f_{bb}\sim 5.1 \times 10^{-13}\, \mbox{erg s}^{-1}\,\mbox{cm}^{-2}$.
  Figure \ref{0437_bb_asca}a shows the ROSAT PSPC data fitted with a double black-body 
  spectrum in which the parameters of the second (harder) spectral component have been  
  fixed according to the ASCA results. The parameters of the soft black-body component 
  have been fitted and imply a temperature of $1.2\times 10^6$ K, $R_{bb}\sim 510\,
  \mbox{m}\,(d/180\,\mbox{pc})$ and a bolometric flux of $f_{bb} = 8.6 \times 10^{-13}\,
  \mbox{erg s}^{-1}\,\mbox{cm}^{-2}$. The column density in this model is fitted to be 
  zero. Due to the small radii both black-body components are attributed to the pulsar's 
  polar cap. The X-ray flux detected with ASCA above 0.7 keV is not pulsed. The authors 
  give a pulsed fraction upper limit of 70\%. 

  An alternative to the double black-body model, also suggested by KTS98, is that 
  the simple power-law component seen by ROSAT and EUVE has a spectral break near 
  the upper boundary of the ROSAT band. In fact, the authors found that both a broken 
  power-law model or a power-law model with exponential cut-off can describe the 
  X-ray data from \PSR very well. The ASCA data imply a spectral break (cut-off)
  near $\sim 1.4$ keV with a slope of about $-4$ for the second power-law component. 
  The broken power-law model fitted to the ROSAT data is shown in Fig.~\ref{0437_brpl_asca}b. 
  The slopes of the first and second power-law components are $-2.35 \pm 0.15$ and $-4.5 
  \pm 1.65$, respectively. The column density is found to be $N_H=0.8\times 10^{20}\,
  \mbox{cm}^{-2}$.  

  A number of different spectral models which assume thermal emission from 
  polar hot spots taking into account the influence of a possible neutron 
  star H-, He- or Iron-atmospheres have been tested recently for \PSR by 
  several authors (see \cite{Moham96} 1996; \cite{Pavlov96}1996; 
  \cite{ZavlinPavlov98} 1998).
  All these models yield spectral fits which are in a somewhat better agreement 
  with the data than a simple black-body spectrum but still leave some unmodeled 
  hard excess above 1 keV.
  The polar cap model of \cite{ZavlinPavlov98} (1998) predicts a pulsed fraction 
  of more than 50\% for the emission above 1.1 keV, in contradiction to what is 
  observed with ROSAT (see Sect.~\ref{tempo} and Table \ref{0437_pf}).

\subsection{Temporal emission properties \label{tempo}}

  Because of  the long time gap between the two PSPC observations and the difference 
  in detector type between the HRI and PSPC we have analyzed the available ROSAT data
  from \PSR independently of each other. The results can therefore be regarded as 
  independent measurements. In the following we give a detailed description of the data 
  analysis for the different data sets.

  Quantities such as the number of counts, the background contribution and the
  number of pulsed counts as well as the fraction of pulsed photons for different
  energy ranges are summarized in Table \ref{0437_pf}.\\[-3ex]

  \begin{table}
  \caption[]{Temporal emission properties of PSR J0437-4715} \label{0437_pf}
  \begin{tabular}{c c c c c c}\hline\hline\\[-1.5ex]
    \quad       Data  &    Range    &   Source  &   Bg       &    Pulsed         &     PF              \\
    \quad        set  &    [keV]    &     cts   &   cts      &      cts          &    [\%]             \\\\[-1.5ex]\hline\\[-1.5ex]
                 {}   &  $0.1-2.4$  &      1200 &    24      &  $326\pm 48$      &   $28\, \pm \;\;4$  \\
    \quad       PSPC  &  $0.1-0.6$  & $\,\,$896 &    19      &  $232\pm 38$      &   $27\, \pm \;\;4$  \\\
    \quad $\!\!1992$  &  $0.6-1.1$  & $\,\,$211 & $\,\,\,$3  &  $\,\,\,67\pm 19$ &   $32\, \pm \;\;9$  \\
    \quad       {}    &  $1.1-2.4$  & $\,\,$102 & $\,\,\,$2  &  $\,\,\,16\pm 14$ &   $18\, \pm 14$     \\\\[-1.5ex]\hline\\[-1.5ex]
    \quad        {}   &  $0.1-2.4$  &      2001 &    80      &  $588 \pm 74$      &   $29\, \pm \;\;4$ \\
    \quad       PSPC  &  $0.1-0.6$  &      1594 &    67      &  $410 \pm 59$      &   $27\, \pm \;\;4$ \\\
    \quad $\!\!1994$  &  $0.6-1.1$  & $\,\,$363 &    10      &  $\,\,107\pm 33$   &   $30\, \pm \;\;9$ \\
    \quad       {}    &  $1.1-2.4$  & $\,\,$158 & $\,\,\,8$  &  $\,\,\,27\pm 20$  &   $16\, \pm 14$    \\
    \quad    Boron    &  $0.1-2.4$  & $\,\,$145 & $\,\,\,4$  &  $\,\,\,37 \pm 13$ &   $26\, \pm \;\;9$ \\\\[-1.5ex]\hline\\[-1.5ex]

    \quad  HRI-94$^*$ &  $0.1-2.4$  &    1055   & $\,\,\,$9  &  $300\pm 50$      &   $29\, \pm \;\;5$  \\
    \quad  $\!\!\!\!$HRI-97 &  $0.1-2.4$  & $\,\,\,$985   & 12  &  $256\pm 40$      &   $26\, \pm \;\;4$  \\\\[-1.5ex]\hline\hline\\[-2ex]
 \end{tabular}
 \begin{minipage}{8.8cm}
 {\footnotesize \footnotesize Errors represent the $1\sigma$ confidence range.\newline
  $^*$Based on data taken on 21. and 22.~Aug. 1994}
 \vspace{-8ex}
 \end{minipage}
 \medskip
 \end{table}

\subsubsection{The 1992 PSPC data}

 \begin{figure*}
  \begin{center}
  \begin{picture}(180,200)
   \put(-5,0){{\psfig{figure=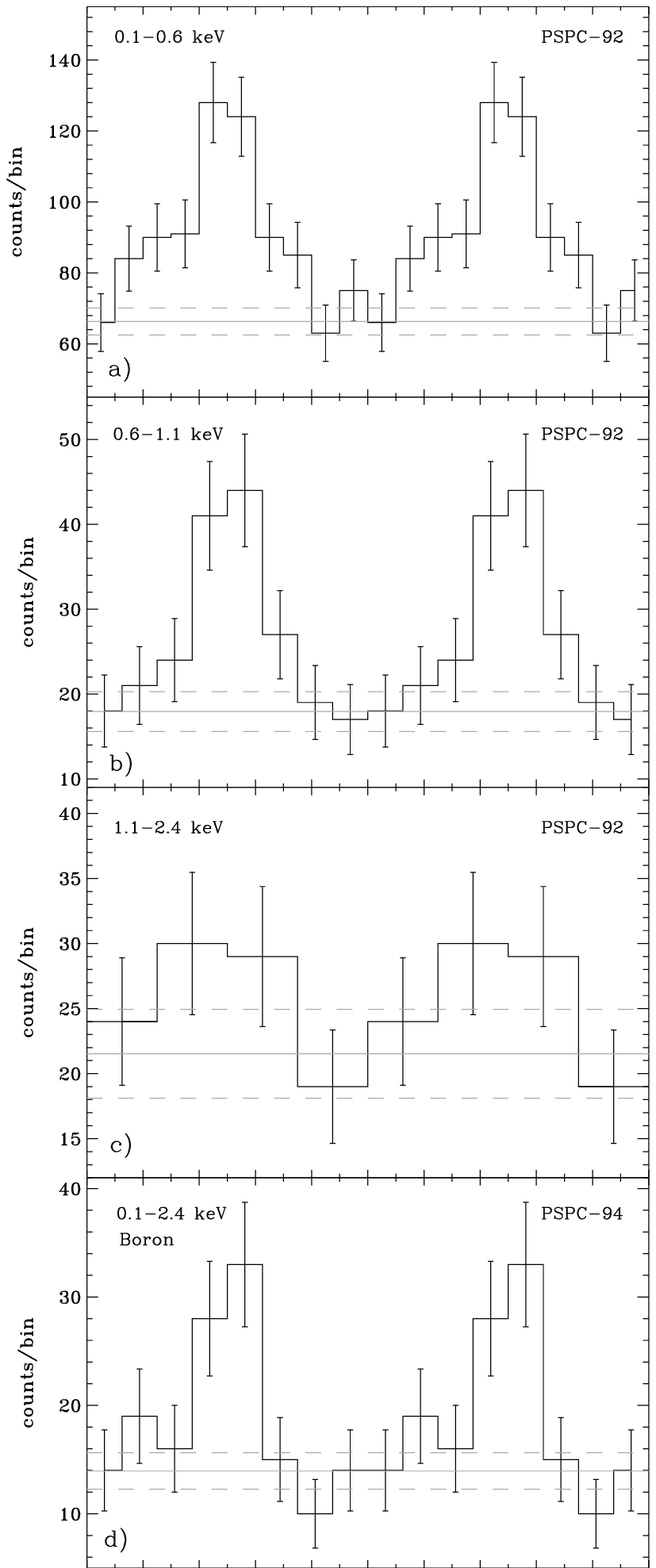,width=9.5cm,height=20cm}}}
   \put(88,0){{\psfig{figure=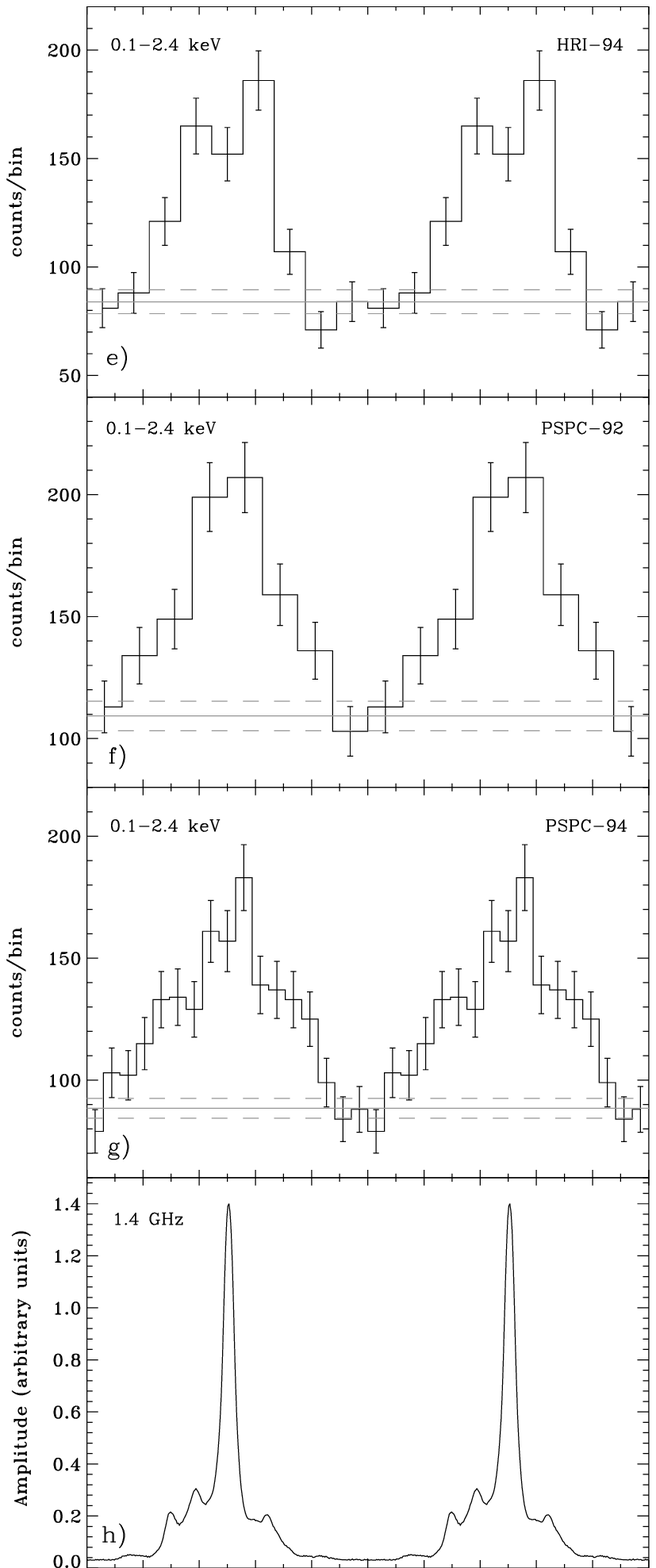,width=9.5cm,height=20cm}}}
   \end{picture}
  \end{center}
  \vspace{0.5cm}
  \caption[]{Integrated pulse profiles of PSR J0437-4715 for different energy ranges.
  Two phase cycles are shown for clarity and $1\sigma$ errors are indicated. Panels
  {\bf a-c} depict the profiles based on the 1992-PSPC data. Panels {\bf d-g}
  show the pulse profiles within the full ROSAT energy range taken with the HRI in
  1994 (e) and the PSPC in 1992 and 1994 (d,f,g), respectively. The DC-level and its
  $1\sigma$ error is indicated by solid and dashed lines, respectively. No phase
  shift is observed for the pulses within the ROSAT energy band. The X-ray pulses are
  found to be consistent with a constant pulsed fraction and a pulse shape which
  appears similar to the radio one shown in panel {\bf h} (see text and Table \ref{0437_pf} 
  for details). The relative phase between the X-ray and radio profiles in this figure 
  has been fixed arbitrarily (see Sect.\ref{0437_rel_phase}).}
  \label{0437_lcs}
 \end{figure*}

  A timing analysis based on this data but using a preliminary pulsar ephemeris led 
  to the discovery of the X-ray pulses from \PSR by \cite{BeckerTrumper93} (1993).
  The results of a reanalysis were presented recently by \cite{HalpernMartinMarshall96}(1996) 
  using an improved ephemeris. However, to perform a timing analysis consistently for all 
  five data sets we have analyzed the PSPC-92 data again using a pulsar ephemeris 
  valid at an epoch close to the ROSAT observational epoch.

  For the photon arrival time analysis of the 1992-PSPC data we have selected 1200 photons 
  from an 70 arcsec aperture centered on the pulsar's X-ray position. The cut-radius was 
  chosen according to the PSPC's point-spread function and contains more than 99.9\%   
  of the pulsar counts. 70 arcsec is large enough to select most of the ultra-soft photons 
  subject to the electronic ghost-imaging effect (\cite{BrielAschenbachHasinger96}1996) 
  which is barely present in this data set. An extraction radius of $\sim 100$ arcsec as 
  used by \cite{HalpernMartinMarshall96}(1996) appears too large and selects an unnecessary 
  large fraction of background photons.  

  The folded pulse profiles for different energy ranges are shown in Fig.~\ref{0437_lcs}a$-$c. 
  The H-Test yields the highest probability for 2 harmonics, 
  indicating a small deviation from a sinusoidal shape. The width of the X-ray pulse 
  is approximately 0.8 rotations ($\approx 290^\circ$), practically the same as observed
  in the radio domain (see Fig.\ref{0437_lcs}h). The fraction of pulsed photons, 
  measured for the different energy ranges by using the bootstrap approach is 
  given in Table \ref{0437_pf}. There is no significant pulsed X-ray emission above 
  1.1 keV in the 1992-PSPC data (see Fig.\ref{0437_lcs}c). Only 102 photons 
  have been detected in this energy range, for which the bootstrap method indicates a
  pulsed fraction of $18 \pm 14\%$. In contrast to our earlier findings (\cite{BeckerTrumper93}
  1993) the bootstrap method indicates that the pulsed fraction is 
  independent of the photon energy in the ROSAT band.

\subsubsection{The 1994 PSPC data}

  In contrast to the 1992-PSPC data, we find that the spatial extent of the pulsar's 
  X-ray counterpart below $\sim 0.25$ keV is strongly dominated by the electronic
  ghost-imaging effect. In order not to exclude these ultra-soft photons from the
  analysis we have selected the pulsar counts below 0.25 keV from an extended region
  chosen according to the expected pattern of photon imaging  
  (\cite{BrielAschenbachHasinger96}1996).
  The PSPC count rate from this area yields a rate of $0.205\pm 0.005$ cts/s, 
  in agreement with the count rate observed in the 1992-PSPC data (\cite{BeckerTrumper93} 1993).
  In total, we have selected 2101 counts for the timing analysis, more than 99\% 
  of the total pulsar contribution. Folding the corrected and barycentered arrival 
  times with the pulsar's rotation frequency resulted in the pulse profile shown 
  in Fig.\ref{0437_lcs}g. The pulse profiles obtained for the energy ranges 
  0.1$-$0.6 keV, 0.6$-$1.1 keV and 1.1$-$2.4 keV are not significantly different 
  from the profiles shown in panel \ref{0437_lcs}a-c. The pulse profile obtained 
  from the observation using the boron filter is shown in Fig.~\ref{0437_lcs}d.

\subsubsection{The 1994 and 1997 HRI data \label{0437_hri_temporal}}

  As already pointed out in Sect.~\ref{0437_obs}, the pulsar's 1994 HRI observation 
  of $\sim 35\,000$ s is distributed over a period of 38 days. Approximately 9800 s 
  of the exposure time was taken between July 18th and August 13th 1994 whereas 
  about 25\,300 s of the exposure was taken within only two days on August 21-22.
  Selecting the photons from an aperture of 10 arcsec radius centered on the 
  pulsar yielded 1484 counts available for the timing analysis. These are 
  approximately 90\% of all pulsar counts and a background contribution of only 
  $\approx 1\%$. 
  Folding the arrival times with the pulsar's rotational frequency (using the same 
  ephemeris as already used for the timing analysis of the 1994-PSPC data) we find 
  from the H-Test that the significance for pulsations is maximized for 3 harmonics 
  at $Z^2_3\approx 91$, indicating a somewhat stronger deviation from a sinusoidal 
  pulse shape than previously found in the PSPC data. However, the fraction of pulsed 
  photons is found to have a rather low value of only $20\pm 4\%$.
  For a consistency check we have therefore performed the same analysis again
  but at this time restricted to the data taken on 1994 August 21.~and 22.~only. 
  The result is interesting in the sense that we find both the significance of the 
  pulsed signal and the fraction of pulsed photons enhanced to $Z^2_3\approx 
  120$ and $29 \pm 5\%$, respectively, although the number of source counts 
  available for the analysis is reduced to 1055. The shape of the pulse profile 
  found in both data sets is not significantly different and shows only a change 
  in the DC level. The profile based on the 21$-$22 August 94 data is shown in 
  Fig.\ref{0437_lcs}e. 

  The differences in significance and pulsed fraction are probably due to the limited 
  stability of the ROSAT on-board clock having a nominal drift of $\approx 8\times 
  10^{-8}$ per day. At the pulsar's observational epoch, clock calibration measurements 
  were only performed once per week. 
  Investigating the residuals of a linear fit from the space craft clock calibration 
  against UTC we find a change in the drift rate approximately in the middle of the 
  first half of the 1994 HRI observation whereas the drift rate stays nearly constant 
  for the remaining time of the observation. 
  From this we conclude that folding all data from the full 38 day observation span 
  results in a smearing of the pulsed emission leading to a reduction of the pulsed 
  fraction. We conclude that the pulsed fraction of $29\pm 5\%$ for the 1994 HRI data 
  is in agreement with the results of the two PSPC observations.

  For the HRI data taken in September 1997 we applied the same analysis as for the 
  1994 data. The results are in agreement with findings deduced from the 1994 HRI 
  observation (see Table \ref{0437_pf}).

\subsubsection{Relative phase between the X-ray and radio pulse \label{0437_rel_phase}}

  In view of the short pulse period of millisecond pulsars and the uncertainty 
  in the ROSAT space craft clock calibration against Coordinated Universal 
  Time (UTC), a comparison of the relative phase between the X-ray and radio 
  pulses for PSR J0437$-$4715 appears difficult.
  Calibration measurements of the satellite's space craft clock against UTC
  are available with a frequency of usually once per week for the pulsar
  observational epochs in 1992 and 1994. A daily clock calibration has been performed
  only since November 1996.

  In order to measure the X-ray pulse arrival relative to the radio pulse we  
  made local fits to the ROSAT clock calibration points against UTC using three 
  and four order polynomial functions. 
  With this approach we find residuals which are in the order of $1-2$ ms.   
  Although this surely represents the limits of the clock accuracy against UTC 
  it still corresponds to a relative error in the X-ray pulse arrival time of 
  approximately $20-30\% (1\sigma)$. In view of these uncertainties the relative
  phase between the radio and X-ray pulse of PSR J0437-4715 cannot be 
  constrained with the available ROSAT data. More accurate timing will be 
  available on AXAF and XMM, which will provide the desired information.  

\section{The globular cluster pulsar  B$1821$-$24$ in M28 \label{1821} }

  PSR B1821$-$24 is an isolated 3.1 ms pulsar in the globular cluster M28.
  Among the few tens of known millisecond pulsars it has the highest spin-down
  energy. X-ray emission from the source was detected with the ROSAT PSPC 
  in March 1991 (\cite{DannerKulkarniThorsett94}1994) and with the ROSAT HRI 
  and the ASCA GIS \& SIS detectors in March 1995 (\cite{DannerKulkarniSaitoKawai97}1997; 
  \cite{SaitoKawaiKamae97}1997). 
  The results of these observations are remarkable in the sense that they
  detect pulsed X-rays in the wide energy band from $0.1 - 10$ keV.

\begin{figure}
  \begin{picture}(80,140)(-6,-12.5)
  \put(0,0){\psfig{figure=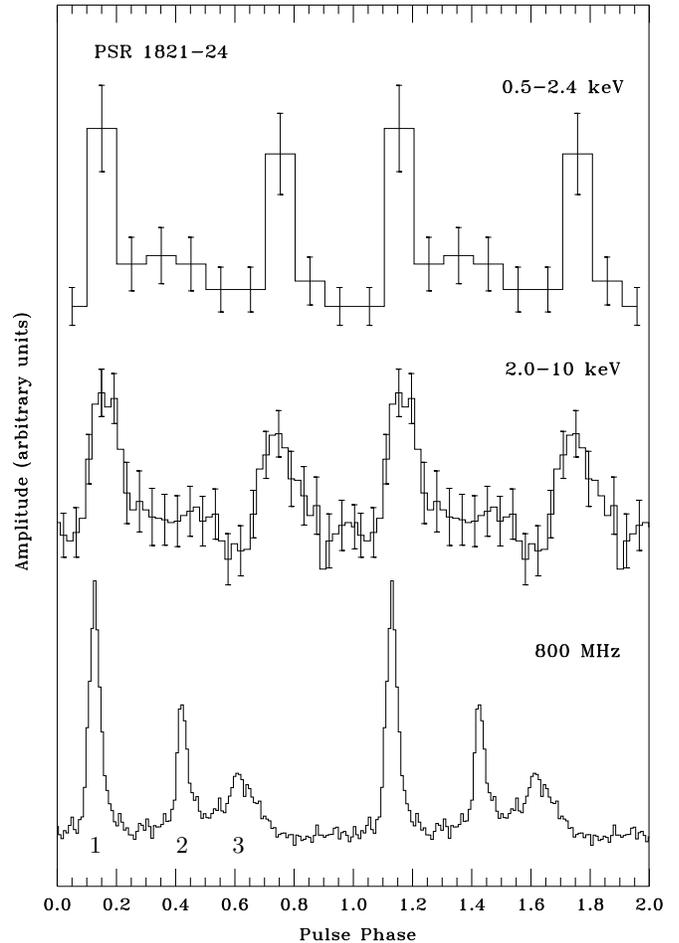,height=13cm,width=8cm}}
  \put(5,13){\small 1}
  \put(16.5,13){\small 2}
  \put(24,13){\small 3}
  \end{picture}
  \vspace{-10ex}
 \caption[]{Integrated pulse profiles of the globular cluster pulsar PSR 1821$-$24
  as observed with the ROSAT HRI (top) by \cite{DannerKulkarniSaitoKawai97}(1997), 
  the ASCA GIS detector by \cite{SaitoKawaiKamae97}(1997) and with the NRAO at 800 
  MHz (bottom) by \cite{BackerSallmen97} (1997). Two phase cycles are shown for 
  clarity. The X-ray pulse profiles are characterized by a double peak structure 
  with a phase separation of $\sim 0.6$ between the two peaks. The radio profile 
  at 800 MHz depicts three pulse components. At this frequency the dominating radio 
  pulse is nearly phase aligned with the primary X-ray pulse. \label{1821_lc}}
 \end{figure}

  Figure \ref{1821_lc} shows the X-ray and radio pulse profile as observed 
  with the ROSAT HRI (top profile), the ASCA GIS-detector (middle profile) and 
  the National Radio Astronomy Observatory in Green Bank (bottom profile). The 
  X-ray light curves are double peaked with a very narrow profile for the pulse
  components. The significance of the X-ray pulses is about $4\sigma$ in the HRI 
  data.

  The high resolution pulse profile observed at 800 MHz was published 
  recently by \cite{BackerSallmen97} (1997). At this frequency, the radio 
  profile shows three pulse components, of which peak \#2 is found 
  to show intensity variations on time scales of hours to days. 
  \cite{BackerSallmen97} (1997) found that the spectral-index of the third 
  and broad radio pulse component is similar to the one observed for 
  the narrow and leading pulse component \#1. The spectral-indices 
  measured between 470 - 1700 MHz are $-1.4 \pm 0.2$ and $-3.1 \pm 0.2$ 
  for the first and second  pulse component (\cite{FosterFairheadBacker91}1991).

  Unfortunately neither the ROSAT nor the ASCA clock calibration is sufficient
  to phase relate the X-ray pulses of a 3.1 ms pulsar with the radio pulse.
  However, recent observations of PSR 1821$-$24 with the Rossi X-ray Timing 
  Explorer (\cite{RotsJahodaMacomb97}1997) have shown that the primary X-ray 
  pulse component appears to be nearly phase aligned with the radio pulse component 
  dominating at 800 MHz. Phase alignment between radio and X-ray pulses is only 
  established (with sufficient accuracy) for the Crab pulsar. Unless the phase 
  alignment is chance coincidence it suggests a common emission site for both the 
  main radio and the dominating X-ray pulse. 

  The spectral analysis based on the ASCA data shows that the pulsars' X-ray 
  spectrum between $0.7-10$ keV is well described by a power-law with a phase 
  averaged photon-index of about $\alpha\approx -2$ (\cite{SaitoKawaiKamae97}1997). 
  The first and the second pulse show photon-indices of $-1.5 \pm 0.3$ and $-2 
  \pm 0.3$, respectively. The spectrum of the globular cluster emission, which 
  is assumed to be the interpulse emission, is not well constrained. The 
  interpulse emission appears to be softer than the pulsar emission, but the 
  data do not allow us to discriminate between several possible spectral models: 
  a power-law with a photon-index of $-2$, an optically thin thermal model 
  with kT=1.4 keV and a kT=6.2 keV bremsstrahlung model yield all acceptable 
  fits.

  Imaging M28 with the ROSAT HRI has shown two separate sources: the point 
  source RX J1824.5-2452P consistent with the millisecond pulsar position and a 
  brighter extended source RX~J1824.5-2452E, whose nature is not yet 
  clear (\cite{DannerKulkarniSaitoKawai97}1997). The presence of the two objects,
  unresolved by ASCA, further confuses the spectral interpretation.
  The power-law spectrum of the interpulse data (i.e.~the spectrum of the
  extended source RX~J1824.5-2452E) favors the model in which RX~J1824.5-2452E
  is a pulsar powered synchrotron nebula, similar to the Crab. 
  An alternative and may be more likely interpretation, however, is that 
  RX~J1824.5-2452E is made of a number of point sources (e.g.~accreting 
  binaries containing white dwarfs or neutron stars) which could not been 
  spatially resolved by the HRI. An interpretation in terms of low accretion 
  LMXBs is also supported from a long term X-ray luminosity study of M28. 
  Our reanalysis of the archival PSPC and HRI data, taken in March 1991 and 
  1995, suggest an X-ray flux variability of M28 on time scales of years. 
  The total ROSAT HRI energy flux taken in March 1995 from M28 is
  about a factor of 3 higher than the total energy flux deduced from the
  March 1991 PSPC observation. This behavior is in line with the results
  recently published by \cite{GotthelfKulkarni97} (1997) who discovered an
  X-ray burst from M28 in the 1995 ASCA data. 

  Estimating the pulsar's energy flux and luminosity from the HRI
  count rate ($3.7 \pm 0.4 \times 10^{-3}$cts/s) yields for the power-law 
  spectrum $f_x=(5.2 \pm 0.6) \times 10^{-13}\; \mbox{erg} \mbox{ s}^{-1}\mbox{cm}^{-2}$ 
  and $L_x = 1.6 \pm 0.2\times 10^{33} (d/5.1\; \mbox{kpc})^2\; \mbox{erg s}^{-1}$ 
  within 0.1$-$2.4 keV, respectively. The latter implies an X-ray efficiency of
  $L_x/\dot{E} \approx 0.8 \times 10^{-3}$. For the interstellar absorption
  we used $N_H=2.8 \times 10^{21}\,\mbox{cm}^{-2}$ which is consistent with the
  ASCA measurements (\cite{SaitoKawaiKamae97}1997) and the column density deduced from
  the dispersion measure.

\section{PSR J0218+4232 \label{0218} } 
  The 2.3 ms pulsar PSR J0218+4232 was discovered by 
  \cite{NavarroBruynFrailKulkarniLyne95}(1995). The pulsar is in a two day binary 
  orbit with a low-mass white dwarf companion and shows significant {\em unpulsed} 
  radio emission throughout the pulse period. The latter has been taken by 
  \cite{NavarroBruynFrailKulkarniLyne95}(1995) as an indication that the magnetic 
  dipole is almost aligned with the rotation axis. The Taylor \& Cordes model 
  indicates a dispersion measure based distance of 5.7 kpc.  

  X-rays from the pulsar were detected on the basis of ROSAT HRI observations 
  by \cite{VerbuntKuiperBelloni96}(1996). Our re-analysis of the archival ROSAT 
  data confirms their findings. Approximately 47 source counts were recorded from 
  PSR J0218+4232 during the 22\,035s HRI observation in August 1995, implying a 
  count rate of $(2.1 \pm 0.4)\times 10^{-3}$ cts/s. This HRI observation was 
  performed $\sim 12$ arcmin off-axis, resulting in a reduced spatial resolution 
  and point source sensitivity. The source is also marginally detected in a 
  serendipitous $\approx 25$ ksec PSPC observation near the edge of the detector's 
  field of view.

 \begin{figure}
 \begin{picture}(88,110)(0,3)
 \put(-8.5,0){\psfig{figure=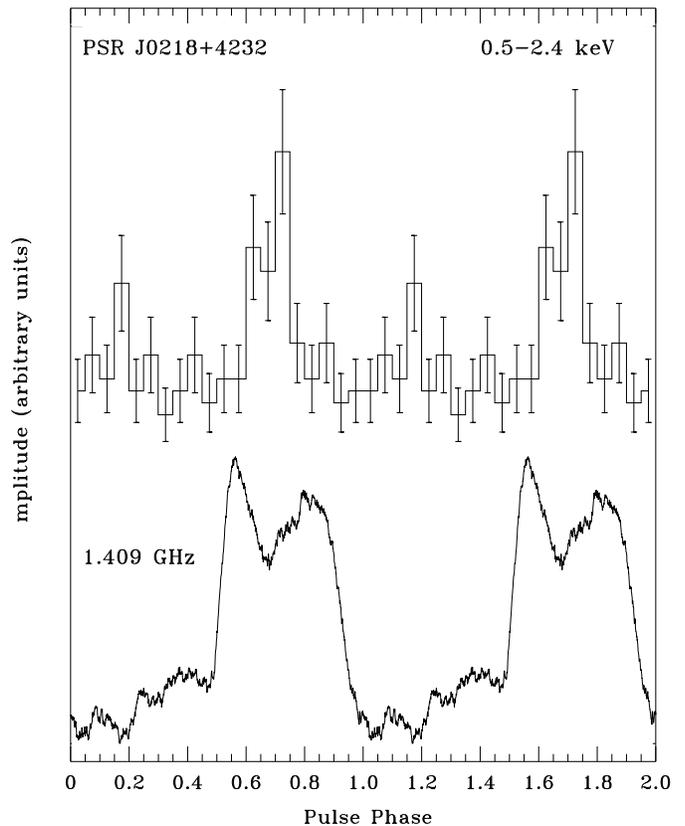,clip=}}
 \end{picture}
 \caption[]{The X-ray and radio pulse profiles of PSR 0218+4232 as observed
  with the ROSAT HRI (top) by \cite{KuiperHermsenVerbuntBelloni98}(1998) and
  with the Effelsberg radio telescope at 1409 MHz (bottom) by Lange et 
  al.~(1998, in prep.). 
  Two phase cycles are shown for clarity. The relative phase between the X-ray 
  and the radio pulse is unknown. In this plot it has been arbitrarily fixed.
  \label{0218_lc}}
 \end{figure}

  X-ray pulses from PSR J0218+4232 have been detected only recently using 
  data from a subsequent ROSAT HRI observation which was performed in July
  1997 (\cite{KuiperHermsenVerbuntBelloni98}1998). About 182 source counts 
  were recorded during an exposure time of 98096 s which revealed the 
  existence of a pulsed signal with a statistical significance of $\sim 
  5\sigma$ and a pulsed fraction of $37 \pm 13\%$. The integrated pulse 
  profiles observed between 0.1-2.4 keV and in the radio channel at 1409 MHz 
  are shown in Fig.~\ref{0218_lc}. As for PSR 2124-3358 (see Sect.~\ref{2124}) 
  there is some indication for the existence of two pulse components in the 
  X-ray lightcurve. 
  \cite{KuiperHermsenVerbuntBelloni98}(1998) mention that the 1998 ROSAT HRI data 
  reveal some 
  marginal evidence for the existence of a compact synchrotron nebula surrounding
  the pulsar. The authors found in their analysis a source extent of about 14 
  arcsec. However, the  HRI point spread function (PSF) assumed by 
  \cite{KuiperHermsenVerbuntBelloni98}(1998) may have been underestimated by 
  neglecting systematic uncertainties 
  in the HRI-PSF due to incomplete attitude correction (see \cite{Boese98} 1998 and 
  discussion therein).     

  No information on the pulsars X-ray spectrum is available so far. 
  The HRI provides no spectral information and the number of counts 
  recorded in the serendipitous PSPC observation foes not allow spectral
  modeling. An estimate of the energy flux and luminosity obtained from 
  the HRI count rate yields  $f_x=1.5 \times 10^{-13}\; \mbox{erg} 
  \mbox{ s}^{-1} \mbox{cm}^{-2}$ and $L_x = 5.7 \times 10^{32}\, 
  (d/5.7\;\mbox{kpc})^2\;\mbox{erg s}^{-1}$ in 0.1$-$2.4 keV, implying 
  an X-ray efficiency of $L_x/\dot{E} \approx 2.3 \times 10^{-3}$. 
  The pulsed energy flux is found to be $f_{xp}^{puls}=(3.9 \pm 1.4) 
  \times 10^{-14}\; \mbox{erg} \mbox{ s}^{-1}\mbox{cm}^{-2}$ 
  (\cite{KuiperHermsenVerbuntBelloni98}1998).  
  The interstellar absorption was adopted from \cite{VerbuntKuiperBelloni96}(1996) 
  who used the maximum color excess $E(B-V)=0.09$ (\cite{BursteinHeiles82} 1982) 
  in the direction of PSR J0218+4232; $A_V=3.1 E(B-V)$ and $N_H=1.79 \times 10^{21}\;
  A_V$ (\cite{PredehlSchmitt95} 1995) which results in $N_H\approx 0.5\times 10^{21}
  \;\mbox{cm}^{-2}$.

\begin{figure}
     \psfig{figure=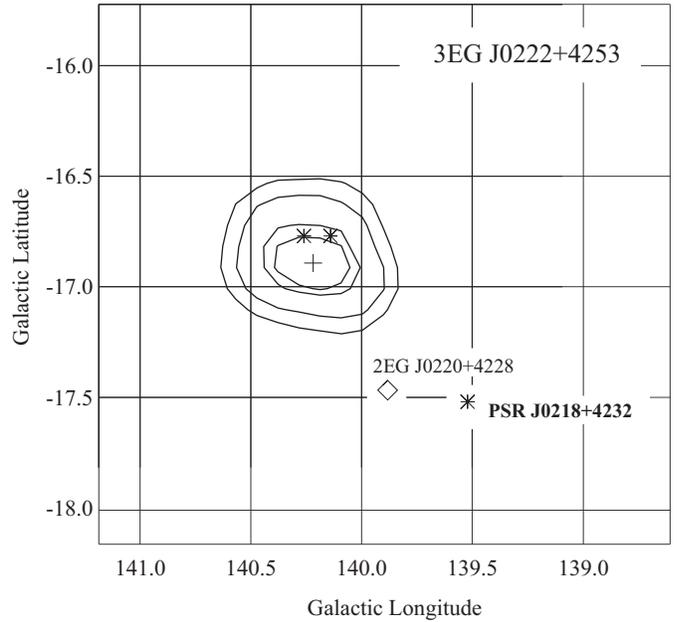,width=8.8cm,clip=} 
  \caption[]{Confidence regions (50\%, 68\%, 95\%, 99\%) of the 
   likelihood test statistics for the position of the EGRET source 
   neighboring the millisecond pulsar PSR J0218+4232. Also shown is 
   the position of the gamma-ray source as listed in the second EGRET 
   catalog (2EG J0220+4228) and the position of the millisecond pulsar. 
   The location of the gamma-ray source is now in agreement with the 
   position of two AGNs, of which 3C66A is found to be the likely 
   counterpart of 3EG J0222+4253 (see \cite{HartmanBertschBloom98}1998 
   for further details). \label{0218gam}}
 \end{figure}

  Although the second EGRET catalog lists a source whose 80\% confidence 
  contour contains the millisecond pulsar (\cite{2EG}1995; 
  \cite{VerbuntKuiperBelloni96}1996), the third EGRET catalog 
  (\cite{HartmanBertschBloom98}1998) which is based on more viewing periods 
  and better data statistics than the 2EG catalog invalidates this putative 
  identification. Figure \ref{0218gam} depicts the likelihood map of the sky 
  region around PSR J0218+4232 based on the third EGRET catalog 
  (\cite{HartmanBertschBloom98}1998). The position of the millisecond pulsar 
  and the gamma-ray source appear well separated from each other. Therefore,
  the millisecond pulsar PSR J0218+4232 is very unlikely the counter part 
  of the neighboring gamma-ray source 3EG J022+4253 which is identified now 
  with the AGN 3C66A (\cite{HartmanBertschBloom98}1998).

\section{The isolated millisecond pulsar J2124$-$3358 \label{2124}}

 PSR J2124$-$3358 belongs to the small group of isolated millisecond
 pulsars. Like PSR J1024$-$0719 and J1744$-$1134, it was discovered by 
 \cite{Bailes97}(1997) during the Parkes 436 MHz survey of the southern sky.
 The pulsar has a rotation period of 4.93 ms and a proper motion corrected 
 period derivative of $\dot{P}=1.077 \times 10^{-20}\;\mbox{s s}^{-1}$,
 implying an upper limit to the pulsar age of $P/2\dot{P}=7.3\times 10^9$
 years and a rotational energy loss of $\dot{E}=3.545\times 10^{33}\;
 \mbox{erg s}^{-1}$. Its close distance of about 250 pc in combination 
 with spin-parameters similar to that of PSR J0437-4715 made the pulsar 
 a promising candidate for X-ray observations.

\begin{figure}
 \begin{picture}(92,80)(0,0)
 \put(0,0){\psfig{figure=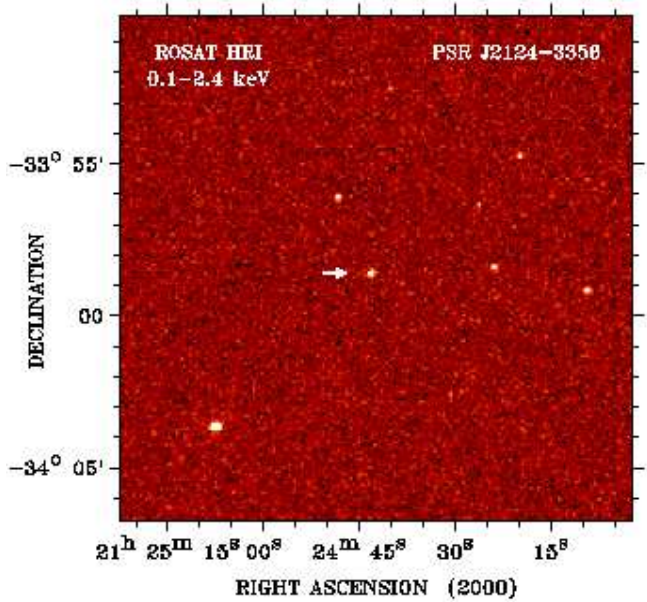,width=8.8cm,clip=}}
 \end{picture}
 \vspace{0ex}
 \caption[]{Sub-image of the HRI's field of view based on the November 1995
  ROSAT observation of PSR J2124$-$3358. The pulsar's X-ray counterpart
  RX J212443.5-335841 is indicated by an arrow. Several neighboring X-ray 
  sources are detected only few arcmin from the pulsar position. 
  \label{2124_subima}}
 \end{figure}

  ROSAT HRI observations of PSR J2124$-$3358 were therefore performed 
  on November 3-7, 1995 with an effective exposure time of 88\,488s.
  Figure \ref{2124_subima} depicts a $16\times 16\;\mbox{arcmin}^2$ sub-image
  of the pulsar field. The pulsar is visible as the X-ray point source 
  RX J212443.5-335841 at the center of the field. A spatial analysis reveals 
  a positional offset between the pulsar's timing position and its X-ray 
  counterpart of $4.8''$, well within the uncertainty of the satellites 
  pointing accuracy. The HRI count rate of PSR J2124$-$3358 is 
  $(2.6\pm 0.2)\times 10^{-3}$ cts/s. 

\begin{figure}
 \begin{picture}(80,120)(-6,-2)
 \put(0,0){\psfig{figure=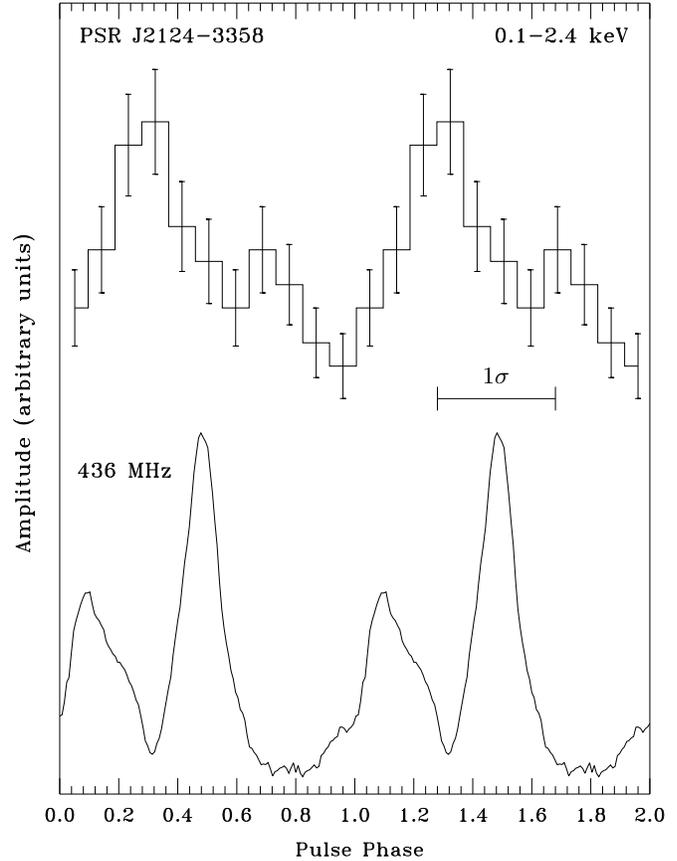}}
 \put(57,64.5){$1\sigma$}
 \end{picture}
 \vspace{0ex}
 \caption[]{Light curves of PSR J2124-3358 as observed with the ROSAT HRI
  in 0.1$-$2.4 keV (top) and the Parkes radio telescope at 436 MHz (bottom).
  Two phase cycles are shown for clarity. Both light curves are characterized
  by two peaks which are separated by $\sim 0.4$ rotations.\label{2124_lc} } 
 \end{figure}

  We applied a timing analysis to 243 photons selected from an $8''$
  aperture centered on the pulsar's X-ray position. This {\em cut} radius includes 
  approximately 97\% of the source photons (\cite{DavidHarndenKearnsZombeck97}1997). 
  About 10\% of the selected photons are background photons.
  After correcting the data for the satellite's motion and applying a barycenter 
  correction, a folding of the photon arrival times implies strong evidence for 
  the existence of a pulsed signal. The resulting light curve is shown in Fig.~\ref{2124_lc}.
  According to the H-Test the pulsations are significant at the $\sim 4\sigma$ 
  level (for two harmonics). This establishes RX J212443.5-335841 as the pulsar's 
  X-ray counterpart and shows the source to be the first galactic solitary \ms 
  X-ray pulsar.
  
  The pulsar's X-ray light curve shows some indication for a double peak as 
  observed in the radio profile at 436 MHz. Both in the radio and the X-ray 
  profile the phase separation between the two peaks appears to be the same. 
  Using the bootstrap method the fraction of pulsed X-ray photons is estimated 
  to be $33 \pm 8\%$.

  For a comparison between the relative phases of the X-ray and radio pulses we have 
  fitted the ROSAT clock against UTC for the days $261-324$ of the year 1995, i.e. 
  for a period of 9 weeks. This is long enough to model the long term clock drift 
  and to cover the observation which was performed during days $307-311$ of 1995. 
  On the other hand we are not limited by clock uncertainties which might have 
  happened long before or after the pulsar observation.   
  Calibration measurements are available close to the pulsar observation on days 
  303 and 310 of 1995. The residuals of our SCC$-$UTC fit were found to be of the 
  order of $\pm 1$ ms, corresponding to a $1\sigma$ error in the relative phase 
  of 20\%. In view of this error it is not possible to draw a quantitative conclusion 
  concerning the phase relation between the X-ray and radio pulse of PSR J2124$-$3358. 
  However, one may speculate that the similarity of both profiles with respect to 
  the phase separation of 0.4 between the pulse components may suggest a phase 
  alignment of the dominant radio pulse with the weak X-ray pulse.

  Assuming a power-law spectrum with a photon-index of $\alpha=-2$ the detected 
  HRI rate corresponds to an energy flux of $f_x \approx 2-3\times 10^{-13}\;
  \mbox{erg} \mbox{ s}^{-1}\mbox{cm}^{-2}$ and an X-ray luminosity of $L_x 
  \approx 1.5-2.2 \times 10^{30} (d/0.25\;\mbox{kpc})^2\; \mbox{erg s}^{-1}$ 
  in 0.1$-$2.4 keV. The latter implies an X-ray efficiency of $L_x/\dot{E} 
  \approx 0.4-0.6 \times 10^{-3}$. The pulsed luminosity is estimated to be 
  $L_{xp}^{puls}=6.2\times 10^{29}\;\mbox{erg s}^{-1}$.
  Based on the radio dispersion measure of $4.6\;\mbox{pc}\;\mbox{cm}^{-3}$   
  (\cite{Bailes97}1997) and the HI in the Galaxy (Dickey \& Lockman 1990) the 
  interstellar absorption is estimated to be in the range $N_H\approx 2-5 \times
  10^{20}\,\mbox{cm}^{-2}$.

\section{Conclusions \label{con_dis} }

  Nine millisecond pulsars are currently detected in the soft X-ray 
  domain. Five of them have been identified only by their positional 
  coincidence with the radio pulsar and in view of their low number 
  of detected counts do not provide much more than a rough flux estimate. 
  These objects are so faint that the high collecting power of AXAF and 
  XMM will be needed to detect enough photons required for a detailed 
  spectral and temporal study in the soft and the hard band beyond 2 keV. 
  More detailed results are found for the other four millisecond pulsars, 
  all of which provide interesting empirical information on the pulsar's   
  X-ray emission mechanisms.  
 
  PSR 1821$-$24 in the globular cluster M28 shows X-ray pulses up to 10 keV. 
  The sharp peaks in the pulse profile and the power-law nature of the spectrum 
  doubtlessly argue for a non-thermal origin of the detected emission. The 
  alignment between the radio and X-ray pulse component adds further support 
  to this interpretation and implies a common emission site for the main X-ray 
  and radio pulse components observed at 800 MHz. For PSR J0218+3242 the sharp 
  peaks found in the X-ray pulse profile also argue for a non-thermal origin 
  of the emission.

  The situation for PSR J2124$-$3358 and J0437-4715 is not that clear. For 
  PSR J2124$-$3358 there is some indication of a double peak structure in 
  the X-ray pulse profile, which in terms of pulsed components and pulse phase
  separation implies a similarity between the X-ray and radio profile
  observed at 436 MHz. However, the significance of the second peak is not
  very strong so that further observations are required to establish this
  similarity.
  Furthermore, for broad pulse profiles the shape of a profile itself is not 
  a strong indicator for the origin of the detected emission: a radiation 
  cone which yields sharp peaks at one aspect angle may well be seen as 
  hardly modulated away from this angle. While sharp peaks indicate 
  non-thermal emission processes the reversal -- soft modulated 
  emission originates from thermal processes -- need not be true.
  The case of PSR J0437-4715 is uncertain as well. Here the bandwidth 
  limitation of ROSAT and the statistical limitations of the ASCA and 
  SAX observations prevent us from conclusively discriminating between 
  multi-component thermal spectra (thermal polar-cap emission) and a 
  non-thermal origin of the radiation.

  Putting the observed emission properties of the detected millisecond 
  pulsars in a somewhat wider frame \cite{BeckerTrumper97} (1997) found 
  recently that the X-ray luminosity of the detected millisecond pulsars
  show the same linear relationship with the same X-ray efficiency as the 
  Crab-like pulsars, indicating that the bulk of their emission is mainly 
  due to non-thermal processes. 
  The  occurrence of power-law spectra in PSR B1821$-$24 and PSR J0437-4715 
  and the similarity between the radio/X-ray pulse profiles seen also for
  PSR J0218+4232 and J2124$-$3358 may be considered as providing additional 
  evidence for a non-thermal origin of the millisecond pulsars' X-ray emission.
  
  In order to  better constrain the origin of the X-radiation of the 
  millisecond pulsars reported in this paper, additional observations 
  with better photon statistics are needed. AXAF and XMM, which 
  are designed to have higher sensitivity and better spectral, timing and 
  spatial performance than ROSAT and ASCA are well suited to study 
  millisecond pulsars and to enhance our understanding of these objects 
  in many aspects.

\begin{acknowledgements}

  The first author wants to thank Okkie de Jager for useful discussions
  on pulse fraction estimates. 
  We also thank Olaf Reimer for the help with the EGRET data and Bob
  Hartman for providing the position of 3EG J0222+4253 prior publication.
  Thanks to Jon Bell, Dick Manchester and Matthew Bailes for the radio
  ephemeris and pulse profiles of PSR J0437$-$4715 and PSR J2124$-$3358,
  to Don Backer and Yoshitaka Saito for the radio and ASCA light curves
  of PSR 1821$-$24 and to Lucien Kuiper for the lightcurve od PSR 
  J0218+4232. Thanks to the Bonner Pulsar Group at MPIfR for supporting 
  us with the 1.4 GHz pulse profile of PSR J0218+4232 prior publication.
  The ROSAT project is supported by the Bundesministerium f\"ur Bildung, 
  Wissenschaft, Forschung und Technologie (BMBW) and the Max-Planck-Society 
  (MPG). We thank our colleagues from the MPE ROSAT group for their support.

\end{acknowledgements}

\end{document}